\documentclass[5p]{elsarticle}

\usepackage{lineno,hyperref,adjustbox,multicol,multicol}
\modulolinenumbers[5]

\journal{Astronomy \& Computing}

\biboptions{authoryear}

\begin{document}

\begin{frontmatter}

\title{Virtual Realities: Is there only one advanced image display that astronomers need?}

\author[1]{Christopher J. Fluke\corref{cor1}}
\ead{cfluke@swin.edu.au}
\cortext[cor1]{Corresponding author}
\author[1]{Hugo K. Walsh}
\author[1]{Lewis de Zoete Grundy}
\author[1]{Brian Brady}

\address[1]{Centre for Astrophysics \& Supercomputing, \\
Swinburne University of Technology, John St, 3122, Australia}

\begin{abstract}
Data visualisation is an essential ingredient of scientific analysis, discovery, and communication. Along with a human (to do the looking) and the data (something to look at), an image display device is a key component of any data visualisation workflow.   For the purpose of this work, standard displays include combinations of laptop displays, peripheral monitors, tablet and smartphone screens, while the main categories of advanced displays are stereoscopic displays, tiled display walls, digital domes, virtual/mixed reality (VR/MR) head-mounted displays, and CAVE/CAVE2-style immersive rooms. 
We present the results of the second Advanced Image Displays for Astronomy (AIDA) survey, advertised to the membership of the Astronomical Society of Australia (ASA) during June-August 2021.  The goal of this survey was to gather background information on the level of awareness and usage of advanced displays in astronomy and astrophysics research. From 17 complete survey responses, sampled from a population of $\sim 750$ ASA members, we infer that: (1) a high proportion of ASA members use standard displays but do not use advanced displays; (2) a moderate proportion have seen a VR/MR HMD, and may also have used one -- but not for research activities; and (3) there is a need for improved knowledge in general about advanced displays, along with relevant software or applications that can target specific science needs. We expect that this is compatible with the experiences of much of the international astronomy and astrophysics research community.  We suggest that VR/MR head-mounted displays have now reached a level of technical maturity such that they could be used to replicate or replace the functionality of most other advanced displays. 
\end{abstract}


\end{frontmatter}

\section{Introduction}
The tools of modern astronomy and astrophysics are many and varied: optical telescopes, radio interferometers, particle detectors, satellites, desktop and high-performance computers, and a variety of mobile computing devices. These tools assist with the collection, processing, storing or sharing of data, with the over-riding purpose of advancing understanding of the Universe. 

Data visualisation plays an important enabling role in the process of turning data into knowledge.  Data visualisation allows the astronomer to view, understand and gain insight into the data they collect or generate.  Data visualisation is an essential ingredient of analysis, and simplifies the process of communicating and sharing results with academic colleagues, students, or the general public.

Within all sub-fields of astronomy, visual inspection and interpretation of data by humans is being steadily augmented or replaced by automated strategies. In particular, this has occurred through: (1) the use of source finders, such as SExtractor for extended sources \citep{Bertin96}, MOPEX for point sources \citep{Makovoz05}, SOFIA for radio spectral line data \citep{Serra15,Westmeier21}, and PyBDSF for astronomical ``blobs'' \citep{Mohan15}; and  (2) 
the growth in application of artificial intelligence (AI) and machine learning (ML) methods.
For general reviews and overviews of AI and ML in astronomy see, for example, \citet{Ball2010}, \citet{Baron19}, \citet{2022arXiv221201493D},  \citet{Fluke20a}, \citet{Longo19}, \citet{2023RPPh...86g6901M}, \citet{Sen22}, \citet{Webb2023}, and \citet{Zelink21}. 

The shift away from `human-only' visual approaches in astronomy has been motivated by the continuing growth in the {\em volume}, or quantity, of data and the {\em velocity}, or rate, at which the data is collected. New classes of instruments, bigger and better telescopes, and improved access to high performance computing, means more data, more often, and a need for faster decision-making and discovery processes in order to generate timely and actionable knowledge.   Indeed, in the future, the vast majority of astronomical data recorded is unlikely to ever be looked at directly by a human.

\subsection{Image display devices}
While astronomers are not planning to eliminate inspecting and interpreting data visualisations from their workflows any time soon -- nor should they -- their ability to complete this task relies on access to a suitable display device.

Along with a human (to do the looking) and the data (something to look at), an image display device is a key component of any data visualisation workflow. The display device plays the crucial role of intermediary between the data and the astronomer's visual system.  In most cases, this is achieved using a fixed-resolution grid of picture elements (pixels), which is either presented directly using a light-emitting screen or via projection onto a surface.

Despite the growth in the total number of astronomical data pixels, voxels, spaxels, points and polygons available for visual inspection,  the majority of current data visualisation activities in astronomy \citep[see][]{Lan21} are completed using {\em standard image display} devices.  This category includes combinations of laptop displays, monitors attached as peripheral displays, or small-scale screens on tablets and smartphones.  

In general, standard displays are used so often because they are: (i) inexpensive;  (ii) readily available; and (iii) transportable. Domain-specific visualisation and analysis software works with minimal additional fine-tuning or set-up.  Laptops, tablets and smartphones are designed to be portable, such that most astronomers are likely to have near-continuous access to a standard display.\footnote{Unless you printed this manuscript on physical paper, it is quite likely that you are currently reading it using a standard display.}

The alternative to the standard display is to use an {\em advanced image display} \citep{Fomalont82,Rots86,Norris94,Fluke06}. In general, the purpose of an advanced display is to present data in a mode that cannot be achieved with a standard display.  The value proposition is that seeing data presented differently may lead to new insights, knowledge or discoveries that would not otherwise be apparent, or would take much longer to obtain, using a standard display. In some cases, this may be because there is a more natural match between the characteristics of the dataset and the display (e.g. viewing all-sky data on a curved surface, real or virtual, which eliminates the distortion inherent when projecting from a spherical coordinate system to a two-dimensional plane).  In others, the potential advantage is the ability to see more data at a time or to encourage collaborative visual analysis to occur.  It is important to reflect that not all astronomers see -- or want to see -- data visualisations presented in the same way \citep{Fluke2023}.

In comparison to standard image displays, advanced displays are often considered to be: (i) expensive; (ii) less readily available; and (iii) constrained to specific locations, which requires the astronomer to leave their office to access and use a suitable visualisation facility.   Additionally, existing analysis software does not necessarily work with advanced displays, and may require a higher level of on-hand technical support.  Consequently, and despite several decades of availability of relevant technologies, the uptake of advanced displays in astronomy and astrophysics research has been very low.\footnote{It is less likely that you are using an advanced display to read this paper, at least in the time period during which this work was authored.} Online surveys, targeting astronomers, provide us with some insight as to why this might be the case.

\subsection{From anecdote to evidence}
\citet{Fluke06} first surveyed the Australian astronomy community in 2005 in order to understand the level of awareness of advanced displays -- the Advanced Image Displays for Astronomy (AIDA 2005) survey.   A request for participants was distributed via e-mail to members of the Astronomical Society of Australia (ASA).  41 responses were received, or about $10\%$ of the ASA membership at the time.  

The AIDA 2005 survey questions were used to gather information on demographics, astronomical instruments and visualization approaches, and the level of awareness of a subset of advanced displays (see Section 4 and the Appendix of \citet{Fluke06} for full details).  Participants were also asked to select from a list of factors, or suggest their own, that were perceived as preventing their use of advanced displays.

A revised and updated version of the AIDA 2005 survey was conducted between June-August 2021 (hereafter AIDA 2021). In particular, we hoped to: (1) investigate whether the level of awareness and/or adoption of several categories of advanced image display technologies had increased; and (2) determine whether the perceived barriers to adoption had changed.  A total of 17 complete responses was obtained through two requests for participation distributed to the membership of the ASA (see Section \ref{sct:advertise}).  

While acknowledging that the number of responses in both the original and updated AIDA surveys was low, we are grateful to those who contributed their time to complete one (or possibly both) of the questionnaires. It is important to note that the 2021 survey was conducted during the second year of the global COVID-19 pandemic, when researchers were trying to operate within the constraints of lockdowns, travel bans, home schooling of children, and complex family and carer arrangements.  This downturn in response rates -- ``survey fatigue'' -- was not unique to the AIDA survey \citep[e.g][]{10.3389,Possami21,Krieger23} during this time period.

\subsection{Overview}
In Section \ref{sct:displays}, we contribute to awareness by providing a brief overview of the most important categories of standard and advanced image displays relevant for use in astronomy and astrophysics research. In Section \ref{sct:survey}, we describe our approach and protocols for undertaking the AIDA 2021 survey, including our advertising strategy, and the corresponding impact on participation.  We present tabular summaries of the responses to the ten survey questions, grouping questions into three categories: (1) usage of standard and advanced image displays (Section \ref{sct:usage}); (2) knowledge of, and interest in, advanced image displays (Section \ref{sct:knowledge}); and (3) benefits and limitations (Section \ref{sct:benefits}).  We encourage the reader to consider what their own responses would have been had they participated in the survey.

In Section \ref{sct:discussion}, we discuss the AIDA 2021 survey results through the use of population proportion confidence intervals and compare key outcomes with the AIDA 2005 survey.  We identify one class of advanced displays -- virtual reality/mixed reality head-mounted displays -- where both awareness and first-hand experience appear to have increased, and explain how these displays can now act as low-cost analogues for other key categories of advanced displays: stereoscopic displays, digital domes, tiled-display walls, and room-scale immersive visualisation environments. 
We present our concluding remarks in Section \ref{sct:conclusions}.

\begin{figure*}[ht]
    \centering
    \includegraphics[width=18cm]{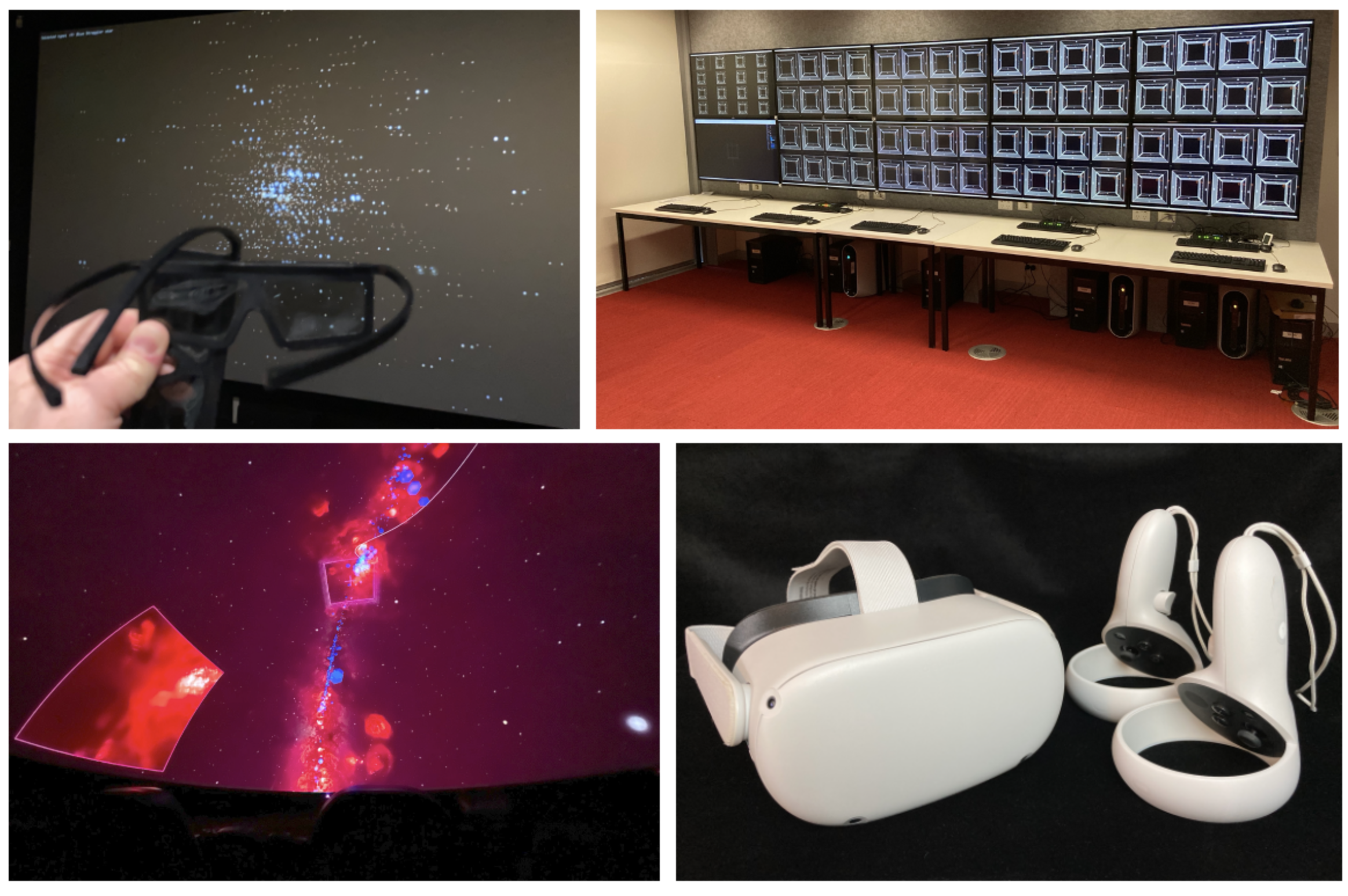}
    \caption{Examples of the four advanced image displays that continue to be the front runners for wider-scale adoption in astronomy and astrophysics research. 
    (Top left) Large-format stereoscopic projection using a two-projector set-up with linear polarising filters and polarisation-preserving silver screen (globular cluster simulation by J. Hurley, visualisation by D. Barnes).
    (Top right)  The Swinburne Discovery Wall is a tiled display wall comprising a matrix of ten 4K UHD displays for a total 84 million pixels, described in \citep{Fluke2023}.
    (Bottom left) Visualisation of the local Milky Way projected onto the dome of the Hayden Planetarium (American Museum of Natural History).
    (Bottom right) The Meta Quest 2 virtual reality head-mounted display with handheld controllers.
    The image of the Hayden Planetarium is courtesy A.Goodman and J.Faherty (see \url{http://milkyway3d.org} for details).  The other images are the work of C.Fluke (CC-BY-NC).  }
    \label{fig:displays}
\end{figure*}

\section{Standard and advanced image displays}
\label{sct:displays}
Most of the standard and advanced imaged displays described in this section share a common origin: with the exception of digital dome projection, they were not designed with the specific requirements of astronomers or astronomical data in mind.  Instead, the displays were general-purpose solutions that can be, and have been, adopted for viewing two-dimensional (2D), three-dimensional (3D) or multi-dimensional data products. This includes pixel-based images, structured volumetric data, or unstructured point/particle clouds \citep{Brunner2002,Hassan11}.  In all cases, a mapping is required from a digital representation of a dataset to a version that can be viewed with a display.

\subsection{Standard image displays}
The two defining features of standard image displays are their pixel dimensions and their physical size.  Selecting a display with more pixels increases the resolution of fine detail in the image, but at a higher total computational cost to generate and display the image.  A larger physical display area (e.g. using a data projector rather than a dedicated computer monitor) can improve the field-of-view or support collaborative inspection, but can also result in visible pixels.  Secondary characteristics, such as the display brightness, screen aspect ratio, and refresh rate, can further impact both the quality and the suitability of the display for displaying different types of data.

\begin{table}
    \caption{Pixel dimensions for a collection of representative standard image displays. FHD is the Full High Definition standard, also referred to as 1080p. WUXGA is the Widescreen Ultra Extended Graphics Array standard. 4K UHD, or 4K $\times$ 2K, is the 4K Ultra High Definition standard. Table columns are the resolution (in pixels) and the total number of pixels (in Megapixels).}
    \label{tbl:pixels}
\begin{adjustbox}{width=\columnwidth,center}
\begin{tabular}{lccc}
{\bf Standard or device}	&	{\bf Dimensions	}		&		{\bf Total}	\\ \hline
FHD	&	1920	$\times$	1080	&	2.07	\\
WUXGA 	&	1920	$\times$	1200	&	2.30	\\
Microsoft Surface Pro 11 (2024)	&	2880	$\times$	1920	&		5.53	\\
Apple iPad Pro (2024) & 2752 $\times$ 2064 & 5.68 \\
14-inch MacBook Pro (2023)	&	3024	$\times$	1964	&		5.94	\\
4K UHD	&	3840	$\times$	2160	&		8.29	\\ \hline
    \end{tabular}
\end{adjustbox}
\end{table}

\begin{table*}
\caption{Characteristics of the key categories of advanced image displays for research data visualisation in astronomy and astrophysics.  Selection of an appropriate advanced display will likely depend on the need for: a display that can be housed in a typical office or at the desktop (Office), high pixel count (Many pixels), stereoscopic 3D visualisation (Stereo 3D), immersion within the dataset(Immersive), physical navigation (Navigable), and suitability for collaborative visualisation and exploration of data (Collaborative). }
\label{tbl:characteristics}
\begin{adjustbox}{width=2.05\columnwidth,center}
    \begin{tabular}{lcccccc}
    {\bf Display type} & {\bf Office} & {\bf Many pixels} & {\bf Stereo 3D} & {\bf Immersive} & {\bf Navigable} & {\bf Collaborative} \\
   \hline 
    Stereoscopic monitor & Yes & No & Yes & No & No & No  \\
    Stereoscopic projection &  No & No & Yes & Partial (A) & Possible (B) & Possible (C) \\
    Tiled display wall & No & Yes & Yes & Possible (D) & Yes & Yes\\
    Digital dome & No & Possible (E) & Possible (F) & Yes & Yes & Yes \\ 
    VR/MR head-mounted display & Yes & No & Yes & Yes & Yes & Possible (G) \\
    CAVE/CAVE2 & No & Yes & Yes & Yes & Yes & Partial (H) \\ \hline
\multicolumn{7}{l}{A: Stereoscopic projection provides a strong sense of immersion, but requires the viewer to look at the projection screen}\\ 
\multicolumn{7}{l}{at all times.}\\
\multicolumn{7}{l}{B: Limited physical navigation is possible within front-projected spaces, improving with rear-projection and head-tracking.}\\ 
\multicolumn{7}{l}{C: Collaborative visualisation is possible when using a large-format projected display, but caveats from note B remain.}\\
\multicolumn{7}{l}{D: Non-stereoscopic tiled display walls can utilise a wide variety of geometrical configurations beyond those of the linear}\\
\multicolumn{7}{l}{matrix shown in Figure \ref{fig:displays}.}\\
\multicolumn{7}{l}{E: Multiple-projector digital dome projection is common, however, there is a requirement for edge-blending between}\\
\multicolumn{7}{l}{projectors and maintenance of uniform brightness.}\\
\multicolumn{7}{l}{F: Stereoscopic dome projection is achievable, however, this is preferable for fixed, forward-facing seating or head-tracking }\\
\multicolumn{7}{l}{of an individual.}\\
\multicolumn{7}{l}{G: Remote collaboration is achievable within VR/MR environments, with opportunities to explore and enhance the use of}\\
\multicolumn{7}{l}{representational avatars.}\\
\multicolumn{7}{l}{H: For correct immersive stereoscopic projection, a single user must be head-tracked, reducing the suitability of the}\\
\multicolumn{7}{l}{ CAVE/CAVE2 as a multi-person environment.}
\end{tabular}
\end{adjustbox}
\end{table*}

The majority of modern standard image displays rely on liquid-crystal display (LCD), light-emitting diode (LED) or organic LED (OLED) technologies, although  legacy \linebreak
cathode-ray tube (CRT) solutions may still be in use in some environments.   Considering a collection of representative current display resolutions (standards or recent releases of specific devices), the typical number of pixels per screen is around $2$ to $6$ Megapixels -- see Table \ref{tbl:pixels}.

For comparison, the Wide Field Channel of the Advanced Camera for Surveys instrument (Hubble Space Telescope) captures images that are 4096 $\times$ 4096 pixels = 16.7 Megapixels in size \citep{Clampin02}, while DECam (Dark Energy Camera; Cerro Tololo Inter-American Observatory) combines 62 individual 4096 $\times$ 2048 charge-coupled devices (CCDs) in a hexagonal configuration for a total of 520 Megapixels per field \citep[e.g.][]{Flaugher06}. In both cases, the pixel sizes of the images exceed the capabilities of most standard displays, which means that a combination of panning and zooming is required in order to completely inspect a full-resolution image.

\subsection{Advanced image displays}
\label{ss:aid}
The advanced image display category includes several important technologies, which we will refer to in the remainder of this work:
\begin{itemize}
    \item Stereoscopic screens and digital three-dimensional (3D) projection environments \citep{Fluke06,Holliman11}, which provide depth perception through the generation of horizontal parallax within a pair of images (Figure \ref{fig:displays}, top left); 
    \item Tiled display walls \citep[TDWs;][]{Sims10,Meade14,Pietriga16,Fluke2023}, which use a matrix of standard monitors to achieve both a much higher pixel count and a larger collaborative workspace (Figure \ref{fig:displays}, top right);
    \item Digital domes \citep{Fluke06,Marchetti18,Jarrett21}, which originated with the hemispherical displays used for public education in the planetarium,  including both large-format shared spaces and smaller-scale vertical domes (Figure \ref{fig:displays}, bottom left);  
    \item Virtual reality (VR) and other mixed reality (MR) or extended reality (XR) head-mounted displays (HMDs), where the user is immersed in a partly- or fully-artificial environment, able to access up to $4 \pi$ steradian of digitally-generated space (see \citet{Deering92} and \citet{Bryson96} for important early work)  (Figure \ref{fig:displays}, bottom right); and
    \item CAVE \citep[CAVE Automatic Virtual Environment;][]{Cruz92} and CAVE2 \citep{Febretti13} room-scale immersive environments, which use either multiple projectors or multiple stereoscopic monitors combined with motion tracking to present a high-fidelity 3D experience.
\end{itemize}

The typical features of advanced image displays are the support for one or more of (see Table \ref{tbl:characteristics}): 
\begin{itemize}
    \item High pixel count (e.g. TDWs and multi-projector digital domes), which can allow an improved match of display pixels to image pixels for very high-resolution imaging (see, for example, Figure 1 of \citep{Meade14});
    \item Stereoscopic or 3D display (e.g. digital 3D projection, VR/MR HMDs, and CAVE/CAVE2), which is particularly suitable for three-dimensional datasets, such as volumetric data from spectral cubes or grid-based simulations, particle simulations, and astronomical surveys (two celestial coordinates and a redshift or distance coordinate); 
    \item Immersive visualisation (e.g. TDWs, digital domes, VR/MR HMDs, CAVE/CAVE2 environments), where the viewer is either placed inside a dataset or more of the viewer's field of view is utilised;
    \item Physical navigation, where the viewer can move around or within the display space to see data from different perspectives; and
    \item Collaborative visualisation, where multiple users can easily view, navigate and explore the same visualisation experience within a shared physical or virtual space.
\end{itemize}

Most advanced displays require the use of either specialised hardware (e.g. VR/MR HMDs) or purposefully-configured environments. Examples here include the curved projection surfaces required for digital domes or the \linebreak
polarisation-preserving silvered screens for digital stereoscopic projection. Other advanced displays can be constructed by using common off-the-shelf components, such as a multiple-monitor tiled displays, which may also need multiple computers with higher-capability graphics cards.  Large format advanced displays often require a dedicated space, which imposes an important constraint: the display is not present in the office for everyday access and use.

Finally, there are advanced displays that have crossed over to become standard displays.   An example here is the 4K ultra-high definition monitor (4K UHD -- see Table \ref{tbl:pixels}), which became widely available due to the closely-related global 4K television market.  Taking the 4K UHD monitor's place as an advanced display is the 8K alternative ($7680 \times 4320$ pixels).  Similarly, ultrawide curved monitors, which share some features with partial digital domes, have become popular within the global computer gaming community.

Astronomers can benefit from the reduction in prices of standard and advanced displays when stable consumer markets take hold.  A drop in customer interest, usually linked to a lack of quality or relevant content, can quickly spell the end of an interesting technology.   Consider the case of stereoscopic televisions and monitors that experienced a rapid growth in consumer interest around 2010-11 followed by a sudden decline, with production by major manufacturers ceasing in 2017.

\section{The AIDA 2021 survey}
\label{sct:survey}  
The AIDA 2021 survey was conducted as part of a larger effort to understand visual discovery in astronomy, with a particular focus on the types of training that astronomers receive (Walsh et al., {\em submitted}). In the present work, we only report on the ten questions regarding standard and advanced image displays.  The project underwent ethical review by, and received approval from, Swinburne University of Technology’s Human Research Ethics Committee (SUHREC) prior to the collection of any survey data.

\subsection{Approach and protocols}
The AIDA 2021 questions were loosely based on those of the 2005 instance \citep{Fluke06}, but with modifications to the age-based demographic categories and the subset of advanced displays considered.  Additional questions were added regarding standard visualisation and analysis configurations, and the use of VR/MR. 

The advanced displays referred to in the AIDA 2005 survey included four options that have retained their relevance, but with minor technical modifications: (1) digital dome projection; (2) multiple-projector tiled displays; (3) stereoscopic projection (single screen); and (4) head-mounted displays (HMDs).  Elsewhere in the present work we: (1) replace multi-projector tiled displays with tiled display walls, as they both aim to provide many more pixels by combining lower-pixel count solutions; (2) generalise stereoscopic projection to stereoscopic 3D displays; and (3) tighten the focus of HMDs to comprise only VR/MR technologies.   

Other displays that were of interest in 2005 are no longer relevant, as they had been superseded or discontinued by 2021: autostereoscopic displays, curved stereoscopic environments, and the eight-wall, rear-projected Virtual Room (see Section 3.2 and Figure 2 of \citep{Fluke06}).  

Examples of all of the standard and advanced displays described in Section \ref{ss:aid} were available in Australia when the AIDA 2021 survey was conducted, however, we did not directly ask respondents where or when they had last seen or used any of these displays. We note that since the survey was conducted, one of the CAVE2 systems in Australia (at Monash University) ceased operation.  Otherwise, apart from some minor updates to specific consumer VR/MR head-mounted displays (e.g. Meta Quest 2 was released in 2020, and the upgraded Meta Quest 3 in 2023), the technology landscape for standard and advanced displays remains largely unchanged since 2021.

As per the approved human research-ethics protocol, participants were provided with access to a consent information statement prior to commencing the survey. Participation in the AIDA 2021 survey was limited to individuals who: (1) were at least 18 years of age; (2) intended to commence, had commenced, or had completed a postgraduate qualification; and (3) were currently undertaking, or intended to undertake, research in astronomy (or Earth observation -- the results presented here only consider respondents with a clear connection to astronomy).  

The survey was administered online using the Qualtrics XM experience management platform.\footnote{\url{https://www.qualtrics.com}}  Data was exported from Qualtrics XM for post-collection analysis and visualisation in Microsoft Excel and R. Analysis was performed using a standard image display configuration comprising a laptop screen plus external monitor (see Section \ref{sct:usage} and Table \ref{tab:q3}).

As a preamble to the AIDA 2021 questions, respondents were presented with the following contextual information: {\em ``The questions in this section relate to your level of knowledge and awareness of various standard and advanced image displays.  An image display refers to any technological device that can be used to enable a visual inspection task. A standard image display refers to a device such as an external monitor, laptop monitor, or similar screen. Such devices are likely to be supplied as the default option to a researcher or student.  An advanced image display refers to a device such as a tiled display wall (> 3 monitors used as a single display space often configured as a matrix of rows and columns), a virtual reality or mixed reality head-mounted display, CAVE-style immersive environments, or a curved surface such as a digital planetarium dome.''}

In Sections \ref{sct:usage}-\ref{sct:benefits}, we group questions into three themes: (1) usage; (2) knowledge and interest level; and (3) benefits and limitations. Questions are discussed in the order in which they were presented to respondents, and have been renumbered here from 1-10.

\begin{figure}[ht]
    \centering
    \includegraphics[width=8.5cm]{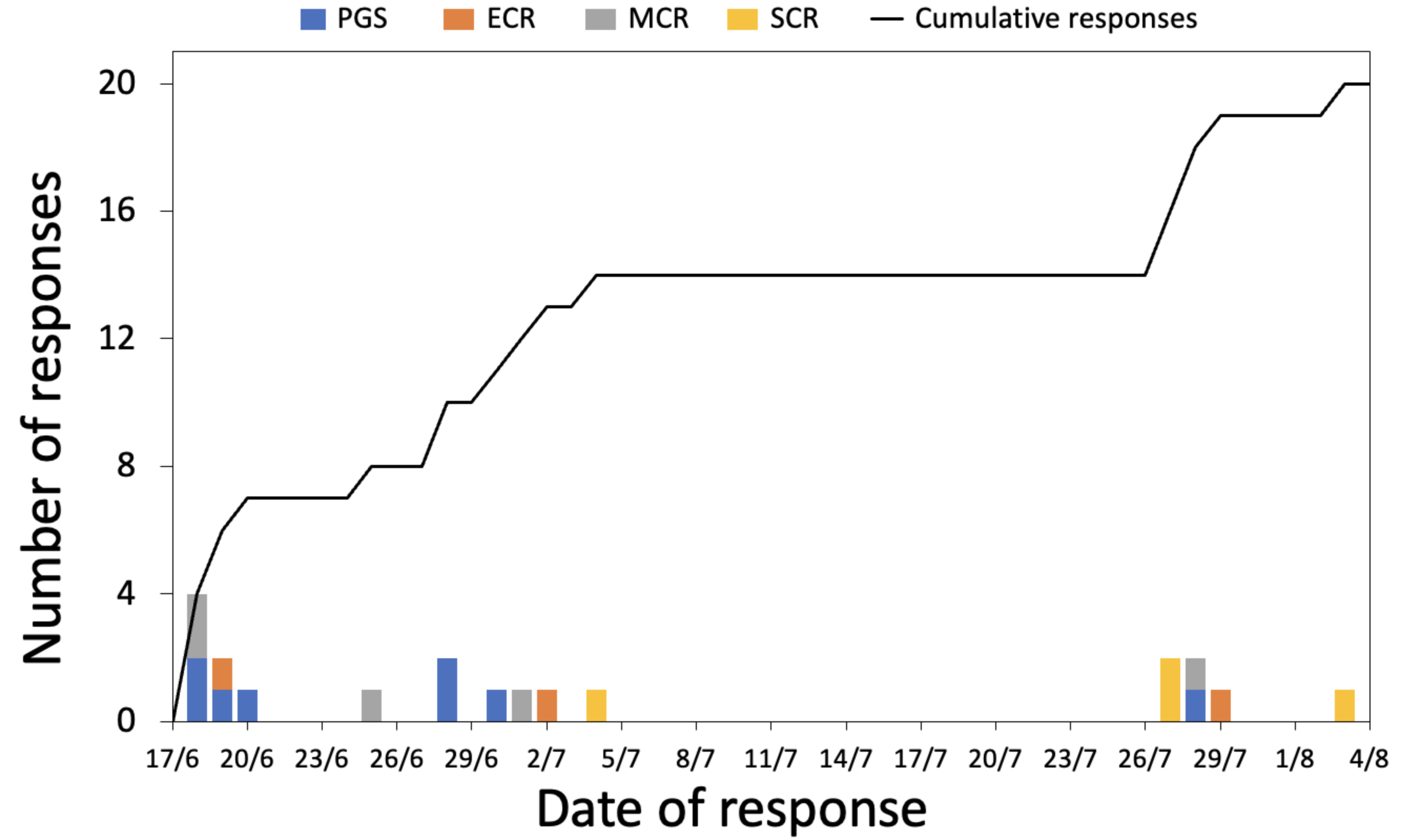}
    \caption{A stacked bar chart showing the number of responses to the AIDA 2021 survey received per day.  The survey was advertised via the ASA e-mail distribution list on 18 June and 27 July 2021.  Respondents are categorised as being postgraduate students (PGS, blue), early-career researchers (ECR, orange), mid-career researchers (MCR, grey), or senior-career researchers (SCR, yellow).  }
    \label{fig:responses}
\end{figure}

\subsection{Advertising and participation}
\label{sct:advertise}
The survey was advertised twice through the ASA e-mail distribution list on 18 June 2021 and 27 July 2021. Responses were collected from 18 June-4 August 2021.  As shown in Figure \ref{fig:responses}, our advertising only elicited 20 responses. At the time of the first advertisement, the number of ASA members was reported as being 729 people, so that AIDA 2021 received responses from less than 3\% of the ASA.\footnote{\url{http://asa.astronomy.org.au/membership/membership-information/}}

From the 20 astronomers who submitted their answers for analysis, three respondents chose not to complete the AIDA-related questions. Their responses are contained in the investigation and analysis of training in visual discovery in Walsh et al. {\em submitted}.

We group our remaining 17 respondents into four career-stage categories: 7 postgraduate students (PGS), 2 early career researchers (ECRs; $<10$ years post PhD), 4 mid career researchers (MCRs; $10-20$ years post PhD), and 4 senior-career researchers (SCR; $>20$ years post PhD).   

There is a clear relationship between the dates of the e-mail advertisements to the ASA and the rate of responses.  This suggests that the main ways to increase the participation rate would have been to: (1) advertise the survey more regularly; and/or (2) use alternative methods that directly target the professional cohort that we wished to understand.

\subsection{Usage of Standard and Advanced Image Displays}
\label{sct:usage}
The first question block in the AIDA 2021 survey comprised four questions regarding:
\begin{enumerate}
    \item[Q1.] Prior usage of standard image displays (see Table \ref{tab:q1});
    \item[Q2.] Prior usage of advanced image displays (see Table \ref{tab:q2});
    \item[Q3.] The combination of displays used most frequently for work-related activities (see Table \ref{tab:q3}); and
    \item[Q4.] The computer operating system used most frequently for work-related activities (see Table \ref{tab:q4}).
\end{enumerate}

\begin{table}[ht]
    \caption{Q1. Have you ever used a standard image display for your research?}
    \label{tab:q1}
    \centering
    \begin{tabular}{cccccc}
{\bf Response} & 
   {\bf PGS} & {\bf ECR} & {\bf MCR} & 
   {\bf SCR} & {\bf Total} \\ \hline
Yes & 6 &  2 & 4 & 4 & 16  \\
 No &  0 & 0 & 0 & 0 & 0  \\
Not sure &  1 & 0 & 0 & 0 & 1 \\ \hline
    \end{tabular}
\end{table}

\begin{table}[ht]
    \caption{Q2. Have you ever used an advanced image display for your research?}
    \label{tab:q2}
    \centering
    \begin{tabular}{cccccc}
  {\bf Response} & 
   {\bf PGS} & {\bf ECR} & {\bf MCR} & 
   {\bf SCR} & {\bf Total} \\ \hline
 Yes & 0 & 0 & 1 & 0 & 1 \\
 No &  5 & 2 & 3 & 4 &  14 \\
 Not sure &  2 & 0 & 0  & 0 & 2  \\ \hline
    \end{tabular}

\end{table}

\begin{table}[ht]
    \caption{Q3. Consider the combination of displays that you use most frequently for completing your work activities.  From the following list of options, please select the components that are the best match to the single display or combination of displays that you use most frequently.  We use the following abbreviations: LS = laptop screen, EM = external monitor, TS = tablet screen, AIO = all-in-one computer.  }
    \label{tab:q3}
        \begin{adjustbox}{width=8.5cm,center}
    \begin{tabular}{cccccc}

{\bf Configuration}	& {\bf PGS} & {\bf ECR} & {\bf MCR} & 
   {\bf SCR} & {\bf Total} \\ \hline
LS	&	0	&	0	&	0	&	1	&	1	\\
LS + EM	&	7	&	1	&	3	&	2	&	13	\\
TS, LS+EM	&	0	&	1	&	0	&	0	&	1	\\
2$\times$EM	&	0	&	0	&	1	&	0	&	1	\\
AIO + 2$\times$EM	&	0	&	0	&	0	&	1	&	1	\\ \hline
    \end{tabular}
    \end{adjustbox}
\end{table}

\begin{table}[ht]
    \caption{Q4. Which one of the following operating systems (OS) do you use most regularly for work-related activities?  The two respondents who did not use a laptop screen (Q3 and Table \ref{tab:q3}) were both Unix/Linux users. }
    \label{tab:q4}
    \centering
    \begin{tabular}{cccccc}
{\bf OS}	&	{\bf PGS} & {\bf ECR} & {\bf MCR} & 
   {\bf SCR} & {\bf Total} \\ \hline
Unix/Linux	&	5	&	0	&	1	&	1	&	7	\\
macOS	&	0	&	2	&	3	&	3	&	8	\\
Windows	&	2	&	0	&	0	&	0	&	2	\\ \hline
    \end{tabular}

\end{table}

\begin{table}[ht]
    \caption{Q5. For each of the Image Displays in the following table, please select the option that is the closest match to your experience with using the Display. Column headings are: NF = Not familiar with the display; NS = Not seen but know what the display is; Seen = Have seen the display, but not used it; UNRP = Use for non-research purposes. The final two columns indicate the regularity of usage of the display for research purposes: $<50\%$ or $>50\%$ of the time. One blank response was noted for the 8K monitor. }
    \label{tab:q5a}
\begin{adjustbox}{width=9.0cm,center}
    \begin{tabular}{ccccccc}
  {\bf Display}       & {\bf NF} & 
  {\bf NS} & {\bf Seen} & {\bf UNRP} & \boldmath$<50\%$ & \boldmath$>50\%$  \\ \hline
     Smartphone    & 0 & 0 & 1 & 15 & 1 & 0  \\
     LCD display & 0 & 0 & 1 & 1 & 1 & 14  \\
     4K monitor & 0 & 1 & 7 & 4 & 2 & 3  \\ \hline
     8K monitor & 0 & 6 & 7 & 1 & 1 & 1  \\
          TDW & 3 & 5 & 7 & 1 & 1 & 0 \\
     Stereo 3D & 3 & 4 & 8 & 2 & 0 & 0\\
     CAVE/CAVE2 & 11 & 3 & 2 & 1 & 0 & 0  \\
     Digital Dome & 2 & 4 & 8 & 2 & 1 & 0 \\ 
          VR/MR HMD & 0 & 3 & 13 & 1 & 0 & 0 \\
          \hline
    \end{tabular}
\end{adjustbox}

\end{table}

\begin{table}[ht]
\caption{A breakdown of the results of Q5 (see Table \ref{tab:q5a}) in terms of the different participant cohorts.  Only the set of six advanced displays is considered.  For each display and cohort, we present a triplet of values collating the three responses: (1) Not familiar with the display; (2) Not seen but know what the display is; and (3) Have seen the display, but not used it.}
\label{tab:q5abreak}
\begin{center}
\begin{tabular}{lcccc}
{\bf Display}	&	{\bf PGS}	&	{\bf ECR}	&	{\bf MCR}	& {\bf	SCR }	\\ 
\hline
8K monitor	&	(0,3,4)	&	(0,2,0)	&	(0,0,1)	&	(0,1,2)	\\
TDW	&	(3,3,0)	&	(0,0,2)	&	(0,1,3)	&	(0,1,2)	\\
Stereo 3D	&	(3,1,2)	&	(0,2,0)	&	(0,0,3)	&	(0,1,3)	\\
CAVE/CAVE2	&	(6,1,0)	&	(2,0,0)	&	(0,2,1)	&	(3,0,1)	\\
Digital Dome	&	(1,3,2)	&	(0,1,1)	&	(0,0,2)	&	(1,0,3)	\\
VR/MR HMD	&	(0,1,5)	&	(0,0,2)	&	(0,0,4)	&	(0,2,2)	\\
\hline
\end{tabular}
\end{center}
\end{table}

Most respondents had used standard displays for research (16/17 = 94\%, with 1 `not sure' response), but had not used an advanced display (14/17 = 82\%).  For Q3, participants were prompted to select from a list of screen options: tablet, smartphone, laptop, all-in-one-computer, single or multiple external monitors, an advanced display or their own choice via a free text option. As none of the respondents selected smartphone, advanced display or used the free text option, those items have been omitted from Table \ref{tab:q3}. The most common combination of displays comprised a laptop screen and an external monitor (13/17 = 76\%).  This suggests that the portability of the computer and one screen was an important factor, but with the need to also access a larger display area for typical work activities.  

Unfortunately, Q3 does not provide insight into who pays for the displays that are used, which may also influence the choice of configurations.  This is more likely to impact PGS and ECR, who may have limited funding available to choose a different compute or display option.  However, as we show in Section \ref{sct:benefits}, the cost of advanced displays is not considered to be a significant impediment to their uptake.

Regarding the operating system used most regularly for work-related activities, Unix-style options (Unix/Linux and macOS) were in the majority (15/17 = 88\%), with a reasonably even split between these two options.  The two non-laptop screen configurations in Table \ref{tab:q4} were both Unix/Linux systems.  

If astronomers are to make use of advanced displays, then the path to adoption is simplified when the displays use the operating system astronomers are most familiar with for research tasks.  As we suggest in Section \ref{sct:virtual}, VR/MR HMDs appear to be growing in relevance, yet they have been more tightly bound to Microsoft Windows environments than other advanced displays.

\subsection{Knowledge of and interest in advanced image displays}
\label{sct:knowledge}
In the second three-question block, participants were prompted to select their level of:
\begin{enumerate}
    \item[Q5.] Experience with a subset of standard and advanced displays (see Table \ref{tab:q5a});
    \item [Q6.] Interest in using specific advanced displays (see Table \ref{tab:q6}); and
    \item [Q7.] Knowledge, interest and ownership of VR/MR HMDs (see Table \ref{tab:q7}).
\end{enumerate}

Although the Smartphone is a ubiquitous device, 15/17 participants responded that they use one only for non-research purposes.  While this was somewhat unexpected, it might be that communication activities related to research (e.g. e-mail, social media, etc) are considered as distinct from data analysis and visual discovery activities. If that is indeed the independent interpretation that each of our survey participants has made, then our confidence increases that the answers provided regarding advanced displays are indeed strongly related to the types of research activities that require a larger-format screen than a Smartphone provides.  The standard LCD display was the most commonly used display: 15/17 used for research purposes, and 14/17 use one more than 50\% of the time.

With the regards to the usage and awareness of advanced displays, four features standout:
\begin{enumerate}
    \item There is limited use of advanced displays.  While 17 responses is insufficient to draw strong conclusions about the broader level of up-take, this result is consistent with experience;
    \item The CAVE/CAVE2 was the one display category that respondents were least familiar with, which is almost certainly linked to the low number of these facilities in Australia -- in part due to the costs for establishment and on-going operations; 
    \item There is limited availability or accessibility of advanced displays.  Aside from VR/MR HMDs, between 6 (8K monitor and digital dome) to 14 (CAVE/ CAVE2) respondents reporting either no familiarity with -- or no prior opportunity to see -- specific advanced displays; and
    \item While there were no instances of use for research purposes amongst the survey participants, all respondents were familiar with VR/MR HMDs, and 13/17 had seen these devices in operation; and
\end{enumerate} 

To further understand awareness and availability, Table \ref{tab:q5abreak} presents a breakdown of a subset of the results of Q5 in terms of the different participant cohorts.  Only the set of six advanced displays is considered.  For each display and cohort, we present a triplet of values collating the three responses: (1) not familiar with the display; (2) not seen but know what the display is; and (3) have seen the display, but not used it. Overall, the PGS cohort was the least aware of advanced displays, either having no familiarity or had not seen specific displays in operation.  However, for the CAVE/CAVE2, this unfamiliarity was evident in 3 of the 4 career stages, suggesting that awareness is not wholly dependent on longevity in the discipline.  

\begin{table}[ht]
    \caption{Q6. For each of the Image Displays in the following table, please select the option that is the closest match to your interest in using the Display in your research. As a metric of overall interest, the average of each column has been calculated and reported in the final row as a percentage.  }
    \label{tab:q6}
    \begin{adjustbox}{width=9.0cm,center}
    \begin{tabular}{lcccc}
    & {\bf Not} & & {\bf Would like to}& {\bf Know how} \\
{\bf Display}	&	{\bf interested}	& {\bf Not sure} & {\bf know more} & {\bf to use}	\\ \hline
8K monitor	& 5 & 6 & 4 & 2 \\
TDW	& 5 & 5 & 5 & 2 \\
Stereo 3D	& 2 & 8 & 7 & 0 \\
CAVE/CAVE2	& 2 & 7 & 7 & 1 \\
Digital Dome & 3 & 9 & 4 & 1\\
VR/MR HMD & 3 & 7 & 6 & 1 \\
\hline
Average & 20\% & 41\% & 32\% & 7\% 
    \end{tabular}
    \end{adjustbox}
\end{table}

\begin{table}[ht]
    \caption{Q7. Low-cost consumer-grade Virtual Reality (VR) and Mixed Reality (MR) head-mounted displays are now available from multiple commercial vendors.  From the following list, please choose all options that apply to your knowledge, interest, and  ownership of VR/MR head-mounted displays.    }
    \label{tab:q7}

        \begin{adjustbox}{width=9.0cm,center}
    \begin{tabular}{lccccc}
{\bf Option} & {\bf PGS} & {\bf ECR} & {\bf MCR} & {\bf SCR}  & {\bf Total} \\ \hline
Never used & 2 & 0 & 1 & 2 & 5 \\
Have used & 5 & 2 & 3 & 2 & 12\\ \hline
Own or can access wireless HMD & 2 & 1 & 0 & 0 & 3\\
Own or can access wired HMD & 1 & 0 & 0 & 0  & 1\\
Have used for entertainment & 4 & 1 & 2 & 1 & 8\\
Have viewed other's data & 0 & 1 & 1 & 1 & 3\\
Have viewed own data  & 0 & 0 & 1 & 0 & 1\\ \hline
    \end{tabular}
\end{adjustbox}
\end{table}

Even though advanced displays, in various forms, have been available for many decades, the results in Table \ref{tab:q6} suggest that there is still limited awareness of what role they can play in supporting research activities in astronomy.  For each advanced display category, no more than 2 respondents knew how to use the display for their research.   While between 2-5 respondents were not interested in specific displays, the message is clear (and consistent with anecdotal experiences of the authors) that astronomers are not sure whether an advanced display can be of benefit.   As a metric of overall interest, the average number of answers for each of the four response categories (not interested, not sure, would like to know more, know how to use) was calculated, with $\sim30\%$ indicating that they would like to know more about the advanced displays.

The third question in this block explored the level of knowledge, interest, and access to VR/MR HMDs.   As was noted above, awareness of VR/MR HMDs was high amongst our respondents, although we identify an inconsistency in the responses to questions Q5 and Q6.  In Table \ref{tab:q6}, 13 participants reporting having seen a VR/MR HMD, and one had used for non-research purposes.  In Table \ref{tab:q7}, 12 astronomers reported that they had used a VR/MR HMD.  This is potentially due to the regularity of use -- a one-off experience in using a VR/MR HMD (i.e. the device had been seen) for multiple survey participants versus more regular usage by one of the participants.  

As we see in the responses to the questions regarding ownership or access (we did not distinguish between these two options), few of the respondents own or can access an HMD.   Use for entertainment purposes (8/17) occured more frequently than use for viewing data (3 viewing other's data, 1 viewing their own data).

\subsection{Benefits and limitations}
\label{sct:benefits}
The final question block examined perceptions regarding both the benefits and the limitations of utilising advanced displays. These questions asked:
\begin{enumerate}
    \item[Q8.] Whether participants saw a potential benefit from using advanced image displays for research (see Table \ref{tab:q8});
    \item [Q9.] If there were particular factors preventing astronomers from using advanced displays (see Table \ref{tab:q9}); and
    \item [Q10.] Which types of interventions, selected from a list or freely posed, would likely have the the greatest impact on access or use of advanced displays (see Table \ref{tab:q10}).
\end{enumerate}

\begin{table}[ht]
    \caption{Q8. In general, do you see a potential benefit from using Advanced Image Displays for your research?}    \label{tab:q8}
    \centering
    \begin{tabular}{lccccc}
{\bf Option} & {\bf PGS} & {\bf ECR} & {\bf MCR} & {\bf SCR}  & {\bf Total} \\ \hline
Yes	&	1	&	1	&	2	&	2	&	6 (35\%)	\\
Maybe	&	1	&	0	&	1	&	1	&	3 (18\%)	\\
No	&	1	&	1	&	1	&	1	&	4 (24\%)	\\
Not sure	&	4	&	0	&	0	&	0	&	4 (24\%)	\\
\hline
    \end{tabular}
\end{table}

\begingroup
\setlength{\tabcolsep}{4pt} 
\begin{table}[ht]
    \caption{Q9. Which, if any, of the following factors do you believe are preventing you or your colleagues from using Advanced Image Displays in your discipline? Please choose all options that apply.}
    \label{tab:q9}

        \begin{adjustbox}{width=9.0cm,center}
    \begin{tabular}{lccccc}

{\bf Reason} & {\bf PGS} & {\bf ECR} & {\bf MCR} & {\bf SCR}  & {\bf Total} \\ \hline
Lack of: \\
$\dots$An application that suits my needs	&	5	&	2	&	3	&	1	&	11	\\
$\dots$Knowledge of available displays &	3	&	2	&	2	&	2	&	9	\\
$\dots$Access to or availability of displays	&	3	&	1	&	2	&	2	&	8	\\
$\dots$Time to develop suitable applications	&	3	&	2	&	1	&	0	&	6	\\
$\dots$Knowledge of how to use	&	2	&	1	&	0	&	2	&	5	\\
$\dots$Time to learn how to use	&	2	&	1	&	0	&	2	&	5	\\
$\dots$Technical support	&	2	&	1	&	0	&	0	&	3	\\
Cost of advanced displays	&	1	&	1	&	0	&	1	&	3	\\
Other: Lack of relevance to personal	&	1	&	0	&	1	&	1	&	3	\\
\hspace{0.3cm} research activities \\
\hline
    \end{tabular}
\end{adjustbox}
\end{table}
\endgroup

\begin{table}[ht]
    \caption{Q10. Considering your own knowledge and interest in the use of Advanced Image Displays, select up to 3 of the following options that would likely have the greatest impact on your ability to access or use Advanced Image Displays to improve your research activities.}
    \label{tab:q10}

        \begin{adjustbox}{width=9.0cm,center}
    \begin{tabular}{lccccc}
{\bf Option} & {\bf PGS} & {\bf ECR} & {\bf MCR} & {\bf SCR}  & {\bf Total} \\ \hline
Improved knowledge	&	3	&	1	&	3	&	2	&	9	\\
Software/application availability	&	3	&	1	&	2	&	1	&	7	\\
Improved access	&	1	&	1	&	2	&	2	&	6	\\
Discipline-specific training &	4	&	0	&	0	&	2	&	6	\\
Simpler development process	&	2	&	1	&	1	&	0	&	4	\\
Availability of technical support	&	1	&	1	&	1	&	0	&	3	\\
Availability of generic training	&	1	&	0	&	0	&	1	&	2	\\
\hline
    \end{tabular}
\end{adjustbox}
\end{table}

24\% of respondents did not see a potential benefit from using advanced displays for their research (Table \ref{tab:q8}).  As we did not explore specific research activities for individuals, for example through case studies, interviews, or other user-centred design approaches, then it is reasonable that advanced displays are not a unique panacea.  The people who are best placed to understand what they currently can achieve -- or are prevented from achieving with their existing visual discovery workflows -- are the astronomers themselves. What we do take encouragement from is that 35\% did see a benefit, with the remainder uncommitted, indicating that there is an opportunity for a growth in adoption. 

For question Q9, participants were presented with a list of potential limiting factors or barriers, based on the outcomes of the AIDA 2005 survey.  A free text option was also provided, with relevance of the displays appearing as the only other clear barrier.  

Lack of access to advanced displays, lack of knowledge of the types of advanced displays that are available, and lack of appropriate applications that suit the specific needs of individuals were the most commonly-selected limiting factors (Table \ref{tab:q9}).  

\begin{table*}[ht]
\caption{Population proportion confidence intervals (CIs) calculated for 90\% confidence level with sample proportion $P_{\rm sp}$, $N_{\rm sample} = 17$ and $N_{\rm population} = 750$. For each of the 10 survey question, we calculate CIs for the most common response(s), compared with all other responses combined.  The population considered here is the membership of the Astronomical Society of Australia. For Q1, the upper confidence level is clipped at 100, with a value of 103.4 calculated. }
\label{tab:ci}
\begin{tabular}{clcccc}
{\bf Q$\#$} & {\bf Question} & {\bf Response} & $N_{\rm response}$ &  $P_{\rm sp}$ &  {\bf CI}\\ \hline
  1 & Have you used a standard image display? & Yes & 16 & 94\% & (	84.6	,	100.0	) \\
  2 & Have you used an advanced image display? & No & 14 & 82\% &(	66.8	,	97.2	) \\
  3 & Which display combination do you use most often? & LS + EM & 13 & 76\% & (	59.2	,	92.9	) \\
  4 & Which operating system do you use most often? & Unix/Linux & 7 & 41\% & (	21.6	,	60.4	) \\
   &  & macOS & 8 & 47\% & (	27.3	,	66.7	)\\
  5  & Which display do you use most regularly? & LCD & 14 & 82\% & (	66.8	,	97.2	)\\
    & Have you seen VR/MR HMD? &  Yes & 13 & 76\% &  (	59.2	,	92.9	)\\
  6 & Which display would you like to know more about? & Stereo 3D & 7 & 41\%  & (	21.6	,	60.4	)\\
  & & CAVE/CAVE2 & 7 & 41\% &(	21.6	,	60.4	)\\
  7 & Have you used a VR/MR HMD? & Yes & 12 & 71\% & (	53.1	,	88.9	) \\
  8 & Do you see a benefit in using Advanced Displays? & Yes & 6   & 35\% &  (	16.2	,	53.8	) \\
  & & No & 4 & 24\% & (  7.5 , 41.0 ) \\
  9 & What prevents you from using Advanced Displays? & Lack of application & 11  & 65\% &(	46.2	,	83.8	) \\
  & & Lack of knowledge & 9 & 53\% &(	33.3	,	72.7	) \\
  10 & Which factor would have the greatest impact? & Improved knowledge & 9 & 53\% & (	33.3	,	72.7	)\\
  \hline
\end{tabular}
\end{table*}

Addressing awareness can be achieved, in part, through works such as this, but a more visible campaign to raise knowledge may be warranted.  Indeed, improved knowledge was identified as the intervention that would likely have the greatest impact on the use of advanced displays to improve research activities (Table \ref{tab:q10}).

Lack of access is more difficult to overcome, as we are not able to influence hardware purchasing decisions -- although we do note that cost was not considered such an important factor as was the case in AIDA 2005 (41\% identifying cost as a barrier).

\section{Discussion}
Despite the small numbers, the AIDA 2021 survey data presents a snapshot of the Australian astronomical research community at a specific point in time.  A lack of responses to the survey may be due to a lack of knowledge of, or engagement with, the topic -- which is an equally interesting and relevant outcome.

We choose to take a pragmatic view that the responses we have obtained are relevant for understanding the broader astronomical research community. If experiences exist within our participant cohort, then they are present worldwide. Even if there is only a small proportion of the global astronomy research community who would like to make an informed decision about the potential benefits of using advanced displays, then it is worthwhile understanding, assessing and ultimately addressing their questions and concerns.   

\label{sct:discussion}
\subsection{Population proportion confidence intervals}
As we have acknowledged, the low number of participants in the AIDA 2021 survey limits our ability to draw conclusions about the global astronomy and astrophysics research communities.  However, if we consider the ASA as our population of interest, we can calculate a population proportion confidence interval to assess the likelihood that the survey responses are representative of the ASA membership.  As the questions were not known to respondents prior to them commencing the survey, the most likely participation bias is an underlying interest in the use of advanced image displays. 

We use the online Select Statistical Services\footnote{\url{https://select-statistics.co.uk/calculators/confidence-interval-calculator-population-proportion/}} calculator to determine population proportion confidence intervals from our sample, calculated for 90\% confidence level (see Appendix A for details of the calculation). We have a sample size of $N_{\rm sample} = 17$ from a population of 729 ASA members (as of June 2021), although we use a slightly higher $N_{\rm population} = 750$ for the calculations.  For each question, we consider the most common response compared with all other responses combined to obtain the (percentage) sample proportion, $P_{\rm sp}$. 

Population proportion confidence intervals (CIs) are presented in Table \ref{tab:ci}, from which we conclude that amongst the ASA membership: 
\begin{enumerate}
\item A high proportion (Q1: CI = 85-100\%) use standard displays -- most likely incorporating an LCD screen (Q3 and Q5: CI = 59-93\% for the laptop screen plus external monitor combination and CI = 67-93\% for use of an LCD) -- but do not use advanced displays (Q2: CI = 67-97\%); 
\item A moderate proportion have seen a VR/MR HMD (Q5: CI = 59-93\%), and may also have used one (Q7: CI = 53-89\%) -- but not for research activities; and 
\item There is a need for improved knowledge about advanced displays (Q10: CI = 33-73\% as the factor with the greatest impact), along with relevant applications that can target specific science needs (Q9: CI = 46-84\%).
\end{enumerate}

With regards to the benefits of advanced displays (Q8), the sample proportions for the definitive Yes (35\%) and No (24\%) responses have the lowest values in Table \ref{tab:ci}. Consequently, the 90\% confidence intervals are that 16-54\% of the ASA membership likely sees a benefit, while 7-41\% see no benefit.  As we have acknowledged in this work (see Section \ref{sct:benefits}), we do not suggest that all astronomers have to see or derive a benefit from advanced displays.  Even if the lower limit of the Yes response represents the true state of the ASA membership, 16\% is still $\sim120$ people in the Australian astronomy community who see benefit, and hence we can expect many more globally.

\begin{table}[ht]
    \caption{Comparing the usage of standard and advanced image displays for research activities between the AIDA 2005 and AIDA 2021 surveys. Responses are reported for Postgraduate students (PGS) and Academic (ACA = ECR+MCR+SCR) cohorts. }
    \label{tab:q12b}
    \begin{adjustbox}{width=9.0cm,center}
    \begin{tabular}{lcccccc}
	{\bf AIDA} & {\bf 2005}	&			&		&	{\bf 2021}	&		&		\\ 
{\bf Cohort}	&{\bf PGS}	&	{\bf ACA}	&	{\bf Total}	&	{\bf PGS}	&	{\bf ACA}	&	{\bf Total}	\\ \hline
Standard	&	88.9\%	&	91.3\%	&	90.2\%	&	85.7\%	&	100.0\%	&	94.1\%	\\
Advanced	&	16.7\%	&	8.7\%	&	12.2\%	&	0.0\%	&	10.0\%	&	5.9\%	\\ \hline
    \end{tabular}
    \end{adjustbox}
\end{table}

\subsection{Comparing the AIDA 2005 and 2021 surveys}
\label{sct:opportunity}
In the 16-year period between the two AIDA iterations, the most significant changes in the availability of advanced displays were the establishment of consumer markets for 4K UHD televisions and monitors, and the emergence of affordable VR/MR hardware from multiple vendors. 

What was not foreseen in 2005, was the widespread adoption of the touch-based tablet and smartphone as ubiquitous portable computing and communication devices.  The public release of the Apple iPhone in 2007 marked the true start of the smartphone revolution/screen-based culture that has impacted how people work with technology for social connectivity, education and work-related activities.

The AIDA 2005 cohort comprised 17 postgraduate students, 10 postdoctoral researchers, 8 academics with permanent positions,  4 contract researchers, 1 undergraduate student and 1 retired academic \citep{Fluke06}.   To simplify analysis, respondents were identified as either students (18 people), equivalent to the PGS cohort, or `seniors' (23 people). We now relabel the non-student category as `academics', noting that this combines the AIDA 2021 ECR, MCR and SCR cohorts.

With regards to usage of standard and advanced image displays for research, by students or academics, there is little change between the results of AIDA 2005 and AIDA 2021: see Table \ref{tab:q12b}.

When asked in AIDA 2005 whether they saw ``{\em a benefit from using advanced image displays for astronomy research''}, 16 astronomers answered `yes' while the remainder selected `perhaps'.  Scepticism was present in the free-text responses to this question, linked to the availability of displays, the time required to learn to use them properly, and a feeling that existing two-dimensional 2D were adequate for visualising 3D datasets. The factors that were perceived to be limiting the uptake of advanced displays were lack of: (1) knowledge (30/41 = 73\% of responses); (2) software tools (19/41 = 46\%); (3) access to local facilities (19/41 = 46\%); and (4) cost of the displays (17/41 = 41\%).

\begin{figure}[ht]
    \centering
    \includegraphics[width=8.5cm]{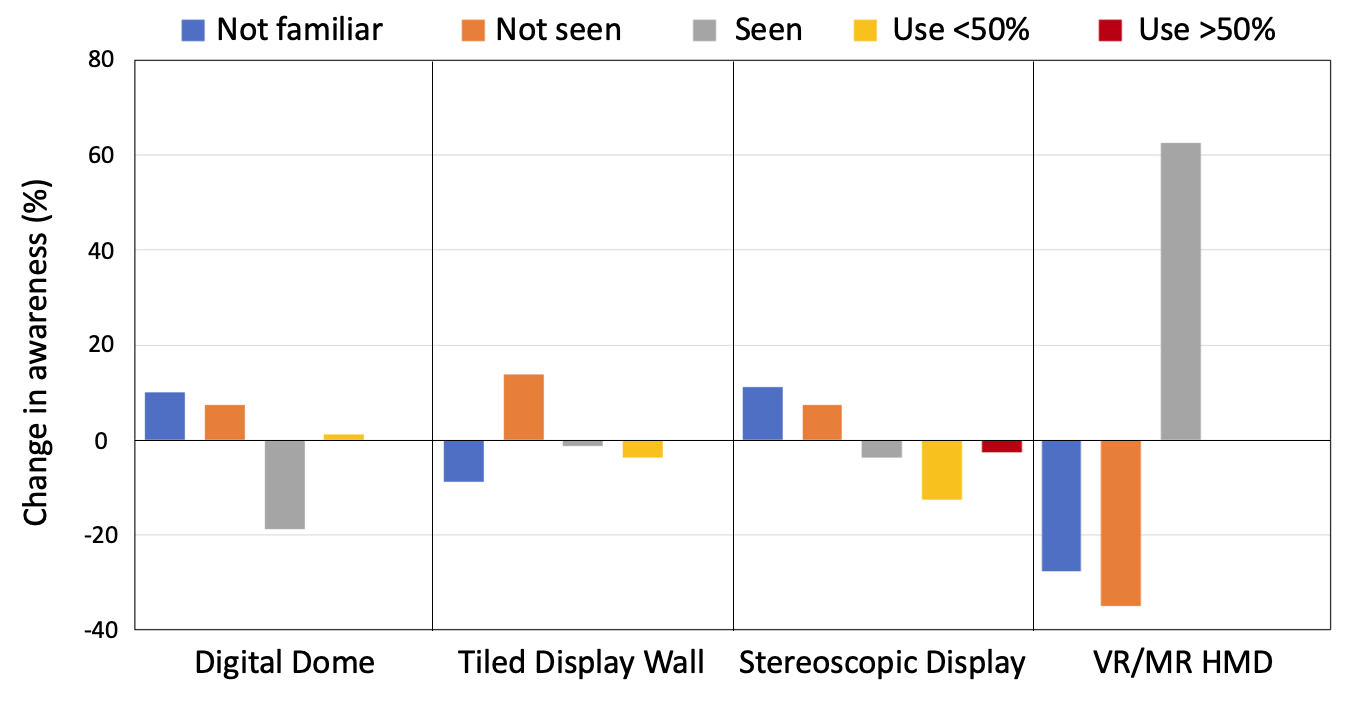}
    \caption{Change in awareness of four advanced displays. For each device category, we determine the difference between the percentage of responses in question Q5 of AIDA 2021 with the equivalent results from AIDA 2005.}
    \label{fig:q5b}
\end{figure}

If we assume that the results of both the 2005 and 2021 AIDA surveys are indeed representative of the experiences of the Australian astronomical community, then several messages emerge:
\begin{enumerate}
    \item Astronomers are still not using advanced displays as part of their typical research workflows -- they use standard displays, with a laptop screen and an attached external (peripheral) monitor proving to be a popular configuration (question Q3).
    \item The level of knowledge regarding the role that advanced displays could play in enhancing visual discovery workflows, and other research activities, has not improved: 73\% of responses in AIDA 2005 and 53\% in AIDA 2021 (questions Q9 and Q10).  
    \item There continues to be a lack of suitable software or applications available, which makes it more difficult to take advantage of advanced displays: 46\% in AIDA 2005 and 65\% (questions Q9 and Q10).  Moreover, Unix/Linux and macOS operating systems are used widely, so for the best return, any software or application needs to be compatible with one or both of these operating systems (question Q4).
    \item Of the four significant categories of advanced displays (Digital domes, TDWs, Stereoscopic 3D, VR/MR HMDs), the only display where there appears to be a growth in awareness, including actually seeing the display in use, is the HMD (question Q5).   We illustrate this in Figure \ref{fig:q5b}: for each device, we determine the difference between the percentage of responses in question Q5 of AIDA 2021 with the equivalent results from AIDA 2005.  The number of users who had seen VR/MR HMDs increased by 50\% compared to 2005, with a corresponding drop in the proportion who were not familiar with or had not seen HMDs.
\end{enumerate}

With these outcomes in mind, we now consider the biggest (current) opportunity to change directions on the adoption and awareness of advanced displays through a strategic focus on a single all-purpose display: the VR/MR head-mounted display.

\subsection{The VR/MR HMD as multi-purpose advanced display}
\label{sct:virtual}
Perhaps the main distinguishing factor between standard displays and advanced displays is their accessibility -- some displays become standard because they are readily accessible to a majority of users (e.g. low-cost laptop screens, tablets, and smartphones), while others remain in the advanced category as they require a dedicated space and are more expensive to both install and operate (e.g. CAVE and CAVE2).   

As a low-cost, transportable device, the VR/MR HMD has the greatest potential to remove all of the access-to-advanced display barriers for researchers.   Moreover, with a growing level of familiarity with VR/MR for non-research purposes, and a significant level of interest through investments by global corporations such as Meta and Apple, there are reasons to remain optimistic that consumer-level VR/MR products will persist as a relevant technology for some time yet.  

Due to improvements in screen resolution, increased frame rates and reduced latency, and usability factors (e.g. wireless streaming to the HMD instead of requirement for tethering to a computer, inclusion of pass-through cameras for safety, enhancement of hand tracking rather than requiring controllers), the digital experience available with current-generation VR/MR HMDs enables them to perform the role of four of the other advanced displays:
\begin{itemize}
\item Stereoscopic display of data works through the simultaneous or sequential presentation of a pair of images with horizontal parallax present.  Controlling what the left and right eye see can can be achieved with (active) shutter glasses, (passive) polarising filters, anaglyph methods, or lenticular screens.  As VR/MR HMDs work by providing separate left-eye/right-eye views within the head-set, they are inherently usable as a stereoscopic display. 
\item In general, astronomers do not keep small- or large-scale domes in their offices, but may be able to obtain access to a suitable dome through collaboration with astronomy education providers.\footnote{For example, the Data2Dome initiative of the International Planetarium Society, see https://data2dome.org and \citet{Marchetti18}} Large-scale digital domes can be aligned with the horizon or tilted to maximise the field-of-view of a forward-facing viewer or audience.  Partial dome-style displays, such as the Cobra Simulation\footnote{\url{https://www.cobrasimulation.com/}}  and Elumenati\footnote{\url{https://www.elumenati.com/}}  products provide a quasi-desktop experience, but with a greater field of view \citep[see][]{Jarrett21}.   The challenge with any dome-based display, as opposed to curved monitor, is the need for one or more projectors that can cover the display area.  For smaller scale domes, this usually requires either a customised fish-eye style lens or a mirror-based solution \citep{Fluke06}, with additional computation required to correctly distort a two-dimensional source image to match the surface shape.  For large domes, the images from multiple projectors must be synchronised and blended. VR/MR HMDs remove both the need for a physical dome surface and a projection solution, allowing the researcher to access a $4\pi$ steradian field of view \citep[see also][for a detailed discussion]{Fluke18}.
\item Tiled display walls are particularly effective when they are used with physical navigation or as part of a collaborative analysis activity \citep{Meade14}.  Being able to step back and see the display in its entirety provides an overall perspective, able to highlight large-scale features or relationships ('the forest').   Then, the viewer can move closer to the display and inspect the final detail of one region ('the trees').  A virtual TDW-style experience can exist within the world-space of virtual reality, where the user is able to walk around a physical space to gain different perspectives on their data. This is possible through a combination of external position tracking (e.g. with infrared beacons placed around the room) or inside-out camera tracking (e.g. with cameras attached to the head-set) and the availability of wireless HMDs that remove the hazards associated with getting tangled in a cable.  Additionally, the introduction of passthrough modes on VR/MR HMDs, such as the Guardian System (for Meta Quest headsets) that is marked out at the beginning of an interaction experience, can reduce trip hazards and collisions with walls or other users sharing the physical space.
\item Addressing the lack of a suitable high-resolution HMD, the original CAVE design used rear-projected screens in a cube or cube-like configuration, combined with head-tracking and significant computational resources \citep{Cruz92}. The CAVE2 was screen-based, using a tiled configuration of stereoscopic monitors, which dramatically increased both the total pixel count and the screen budget \citep{Febretti13}. In both cases, there is a level of redundancy in the graphics pipeline, as content is generated that is not seen by a lone, head-tracked viewer.  VR HMDs provide the actual experience that CAVE/CAVE2 were designed to emulate, but at substantially lower cost, and without the need for a dedicated facility.  
\end{itemize}

For VR/MR HMDs to become a practical and usable advanced display for astronomy, there are barriers and limitations that need to be overcome.  These include factors that were examined in the AIDA 2021 survey, in particular the lack of suitable software applications and the lack of time for researchers to develop suitable applications (question Q9), and incompatibilities with existing operating systems (question Q4). 

A growing assortment of platforms, libraries and software for VR have now been developed, trialled and assessed within astronomy. Recent examples include the use of Unity\footnote{\url{http://unity3d.com}} \citep{2018CAPJ...24...25F,2020EGUGA..2211708T,2020LPI....51.2916V}, A-Frame WebVR\footnote{\url{https://aframe.io/}}\citep{Fluke18}, Universe Sandbox\footnote{\url{http://universesandbox.com}} \citep{2020AAS...23520204S},  PointCloudsVR\footnote{\url{https://github.com/nasa/PointCloudsVR}} \citep{2020AAS...23535704K}, and \linebreak
iDaVIE\footnote{\url{https://github.com/idia-astro/iDaVIE}} for spectral data cube visualisation\citep{Jarrett21}.      

Additionally, vendors are continuously changing technical specifications (e.g. platforms and operating systems that are supported), introducing new features (e.g. integrated hand-tracking for more natural interaction) that can make prior solutions redundant, or ceasing development of products based on the level of general consumer interest (e.g. Google's decision to discontinue support of the Daydream View HMD in October 2019).  These external factors have made it difficult for preferred HMDs, and hence data-to-VR workflows, to be identified.

From a technical stand-point, the current challenges limiting adoption of HMDs for astronomy are: (1) the pixel dimensions of the image presented to each eye -- relevant when considering very high-resolution astronomical images or point-based simulations with billions of particles; (2)  the latency in generating highly-detailed imagery that must be updated for even minor head or body movements in order to avoid motion sickness; and (3) the need to access local high-speed wireless networks (e.g. 5G networks) in order to take advantage of direct-to-headset streaming.  

Considering the growth in academic literature pertaining to VR/MR in astronomy, there is evidence that experimentation and adoption of this advanced image display is gaining momentum. Building on the pioneering work of \citet{7004282}, \citet{Schaaff15}, and \citet{Ferrand16}, amongst others, who all wrestled with vendor-specific Software Development Kits (SDKs) and early-adoption of game engine solutions such as Unity, there are promising signs that VR/MR may now be entering the ``plateau of productivity'' in the Gartner hype cycle.\footnote{\url{https://www.gartner.com/en/research/methodologies/gartner-hype-cycle}}

\subsection{Adoption of virtual reality in astronomy}
\label{sct:limitations}
Several broad application categories have emerged within astronomy and planetary science for VR-based scientific data visualisation and exploratory data analysis:
\begin{itemize}
    \item {\em General-purpose tools for multi-dimensional data exploration}: available as standalone solutions
 \citep{2019zndo...3348451K}, or integrated with the Virtual Observatory \citep{2017ASPC..512..485P}, the  Data and Analysis Centre for Exoplanets \citep{2019ASPC..523...25A}, and for analysis of GAIA data \citep{2019ASPC..523...21R}.  VR has also opened up new modes of immersive multi-sensory data exploration via data sonification \citep{2019IAUS..339..251C}.
 
    \item {\em Targeted exploration and visualisation}: including applications to the quantitative study of the three-dimensional structures supernova remnants  with the
   E0102-VR application \citep{Baracaglia20}, candidate identification and structure analysis of
    circumstellar disks \citep{2020AAS...23535704K},         morphological and physical analysis of giant molecular clouds  \citep{2019MNRAS.484.2089R}, mapping of post-starbust E+A galaxies in rich clusters of galaxies \citep{Liu21a,Liu21b}, 
        tomographic reconstruction of foreground absorption fields from Lyman-$\alpha$ mapping \citep{2018ApJS..237...31L}, visualisation of high-resolution  planetary images \citep{2017EPSC...11..481E}, the discovery of disks in nearby young stellar associations that were found in Gaia data using VR \citep{Higashio22}, and exploration of the membership of stellar clusters \citep{2024AAS...24345816R}. With the COVID-19 pandemic providing additional motivation for remote collaboration, \citep{Milisavljevic21} developed the Collaborative Astronomy VR platform, with an early application in the analysis of supernova remnants.

\item {\em Spectral line cube analysis}:  
iDaVIE was used to distinguish between anomalous gas and the main disc of two spiral galaxies observed in the WALLABY Pilot Data Release 1 \citep{Westmeier22}, contributing to the discovery of two potential polar ring galaxies \citep{Deg23}. Of note was the use of interactive data masking performed within iDaVIE's immersive environment to enhance kinematic modelling of the neutral hydrogen (H{\sc i}) gas distribution -- see also \citet{Kleiner21}. Other use of iDaVIE includes visualisation of H{\sc i}-rich galaxy groups \citep{Glowacki24} and validation of source-finding outputs 
\linebreak\citep{Maccagni24} with spectral line data from MeerKAT \citep{Jonas2016}.

    \item {\em Simulation visualisation}: VR provides alternative ways to assess and examine the outputs of computational simulations, for example, magnetohydrodynamical models from astrophysical simulations \citep{2019sros.confE..76B,2019RNAAS...3..176O},
    hydrodynamic simulations of the Galactic centre \citep{2018cosp...42E2915R,2019HEAD...1711286R}, and
       general-relativistic ray-tracing of accreting supermassive black holes  \citep{2018ComAC...5....1D}.  
     \item {\em Lunar, planetary and small-body surface reconstruction}: the availability of 3D topographical datasets has enabled VR-based Lunar surface simulations for robotic testing  \citep{2018LPICo2063.3095W,2020LPICo2241.5017M}, and investigation of the Martian surface via the 3D Digital Outcrop Model (DOM) obtained from photogrammetry undertaken with the Mars Science Laboratory rover Curiosity 
        \citep{2018EGUGA..2013366L,2019EGUGA..21.3939L,2020P-SS..18204808C,2020EGUGA..22.5714W}. As a component of mission control for synthetic Lunar missions, \citet{2019AGUFM.P33C..08O} demonstrated the potential for VR to provide improved situational awareness for human operators in mission operations. 
 \end{itemize}
 
There has been a corresponding growth in the use of VR for education and new modes of outreach and communication in astronomy, planetary science, and space exploration. With its immersive nature providing a natural way to view all-sky information, VR has been used  successfully to build virtual planetaria \citep{2019EGUGA..2112701A}, including the  American Astronomical Society's WorldWide Telescope \citep{2018ApJS..236...22R}.  \citet{2022CAPJ...31...28I} discuss the creation of a virtual astronomy exhibition, including the importance of considering viewer comfort.  For an overview of the educational opportunities, including perspectives from educators, see \citet{Kersting24}.  The potential for VR to transform the conference experience is now being explored through the Future of Meetings community \citep{2021NatAs...5..213M,2023NatAs...7.1412M}.

In several instances, such as the work by \citet{2018CAPJ...24...25F}, there is a dual use of VR for both outreach and as a tool for scientific visualisation and planning. Examples here include a walk-through of the Cassiopeia A supernova remnant \citep{2018CAPJ...24...17A}, realistic simulations of the WISPR camera of the Parker Solar Probe \citep{2018shin.confE.194S}, and the programs initiated at the Inter-University Institute for Data Intensive Astronomy (IDIA) Visualisation Lab \citep{Marchetti18}.   In this last work, a multi-format approach to data-intensive visualisation utilises the planetarium dome, large format desktop displays, and virtual reality. \citet{2020MNRAS.497.2954L} used these tools to aid identification, and build comprehension, of sub-structures in the 2MASS Redshift Survey  galaxy catalogue \citep{2019ApJS..245....6M}.

\subsection{The Python in the room}
As explained in Section \ref{sct:survey}, the results presented in this work were obtained from responses to 10 questions that were part of a larger survey effort to understand visual discovery in astronomy (Walsh et al., {\em submitted)}.  One question from that full survey is relevant to consider here, as it addresses a missing piece in the adoption-of-advanced-displays puzzle: {\em Which one or more of the following software packages, languages, or tools do you use most regularly to develop computer programs or processing scripts to analyse or visualise data?}  The choices presented included C/C++, Python, R, and Fortran, amongst others, including an ``Other'' free-text option -- see Table \ref{tab:language}.  Responses were obtained from 82 people, comprising 15 PGS, 18 ECRs, 19 MCRs and 30 SCRs.

 There is a general understanding that Python has become the most important and widely-used programming language in astronomy, as evidenced by the significant citation impact of community efforts such as Astropy\footnote{\url{https://astropy.org/)}} \citep{2013A&A...558A..33A} and SciPy\footnote{\url{https://scipy.org/}} \citep{2020NatMe..17..261V}. 
 As anticipated, this outcome was clear in the responses we gathered, with 80\% of the survey participants indicating that they used Python -- twice as many as R, which was the next closest option. 

For astronomers to adopt advanced displays in general, or VR/MR HMDs as multi-purpose displays, there is a clear need to bridge the gap between Python scripting and gaming engines (especially Unity and Unreal Engine\footnote{\url{https://www.unrealengine.com/}}).   Here, another discipline -- vision science -- has demonstrated the required pathway.  The Perception Toolbox for Virtual Reality\footnote{\url{https://ptvr.inria.fr/}} \citep[PTVR][]{10.1167/jov.24.4.19}, which targets HTC Vive Pro HMDs, provides a translation layer between Python and Unity. The workflow encourages the researcher to focus on the experiment they wish to define and conduct, creating Python scripts that the software uses to create the relevant Unity assets and environments.  We note that this is a similar approach to that taken by \citet{Barnes06}, with the  {\sc s2plot} 3D programming library acting as a layer above OpenGL\footnote{\url{https://www.opengl.org/}}, providing straightforward access to multiple types of advanced display.\footnote{Both the stereoscopic and tiled-display wall visualisations in Figure \ref{fig:displays} make use of {\sc s2plot}.}

\begin{table}
\caption{Survey participants were asked to nominate the software packages, languages, or tools they used most regularly to develop computer programs or processing scripts to analyse or visualise data. The 82 respondents selected one or more options from the list in the table, along with a free-text choice of Other.} 
\label{tab:language}
    \begin{tabular}{cccccc}
{\bf Option} & {\bf PGS} & {\bf ECR} & {\bf MCR} & {\bf SCR}  & {\bf Total} \\ \hline
        Python & 14 & 17 & 15 & 20 & 66 \\ 
        R & 4 & 7 & 7 & 15 & 33 \\ 
        Unix shell  & 5 & 3 & 3 & 10 & 21 \\        
        C & 3 & 4 & 5 & 5 & 17 \\ 
        IDL & 3 & 1 & 5 & 8 & 17 \\
        Fortran & 1 & 4 & 3 & 8 & 16 \\ 
        C++ & 2 & 4 & 2 & 0 & 8 \\ 
        Perl/PDL & 0 & 0 & 2 & 1 & 3 \\ 
        MATLAB & 0 & 1 & 0 & 1 & 2 \\ 
        Java & 0 & 1 & 0 & 0 & 1 \\ 
        Other & 2 & 2 & 2 & 8 & 14 \\ \hline    
    \end{tabular}
\end{table}

\section{Conclusions}
\label{sct:conclusions}
Through the AIDA 2021 survey advertised to the membership of the Astronomical Society of Australia, we have obtained a snapshot (at a point in time) of the level of awareness and interest in the use of advanced image displays in astronomy and astrophysics research.  As we have emphasised throughout, our cohort of 17 responses sampled from a population of 750 ASA members does not allow us to draw far-reaching conclusions.  There is, however, statistical relevance at the 90\% confidence level when considering the most frequent answers to several key questions.

We find that astronomers: (1) use standard image displays, especially LCD monitors (67-97\% of ASA members), as a critical component of their research workflows (85-100\% of ASA members); but (2) do not use advanced displays (67-97\% of ASA members).  

Compared to the AIDA 2005 survey, awareness of VR/MR HMDs appears to have improved, with 59-93\% astronomers likely to have seen such a display in action, but have not used one for research purposes.  VR/MR HMDs may be able to play a research role as a portable, multi-purpose advanced display that can support immersion, stereoscopic modes, all-sky or wide field-of-view visualisation without the need for a physical dome surface, and physical navigation at a fraction of a cost of a tiled display wall or CAVE/CAVE2 without requiring a dedicated room.  

Aligned with increased awareness of VR/MR HMDs, there are data exploration tools ready for use (for example, Universe Sandbox, PointCloudsVR and iDaVIE) -- and perhaps more importantly -- scientific discoveries or insights enabled with VR-based visualisation and analysis \citep[e.g.][]{Higashio22}.

The benefit of using advanced displays is still very much in question (only 16-54\% ASA members see a benefit), however, interest does not have to be universal for an alternative solution to be worthy of further exploration. Lack of suitable software applications (46-84\% of ASA members) and lack of knowledge (33-73\%) appear to be the two most significant barriers to experimentation, adoption and use.  Our hope is that this work contributes to addressing the lack of knowledge, while presenting a prompt for some astronomers to think more carefully about the role advanced displays could play in their own research workflows now and into the future.  

\section*{Acknowledgements}
We acknowledge the Wurundjeri People of the Kulin Nation, who are the Traditional Owners of the land on which the research activities (analysis and interpretation) were undertaken. During the period when this work commenced, Christopher Fluke was the SmartSat Cooperative Research Centre (CRC) Professorial Chair of space system real-time data fusion, integration and cognition.  This work has been supported by the SmartSat CRC, whose activities are funded by the Australian Government’s CRC Program.   
Hugo Walsh, Lewis de Zoete Grundy and Brian Brady are supported by Australian Government Research Training Program Scholarships.  CF thanks Alyssa Goodman, Jackie Faherty, and others in the AstroViz Community of Practice for providing images of digital domes.
This work is dedicated to the memory of our friend and colleague, Tom Jarrett, who understood and championed the potential of advanced displays for astronomy.

\bibliography{main}

\begin{thebibliography}{92}
\expandafter\ifx\csname natexlab\endcsname\relax\def\natexlab#1{#1}\fi
\providecommand{\url}[1]{\texttt{#1}}
\providecommand{\href}[2]{#2}
\providecommand{\path}[1]{#1}
\providecommand{\DOIprefix}{doi:}
\providecommand{\ArXivprefix}{arXiv:}
\providecommand{\URLprefix}{URL: }
\providecommand{\Pubmedprefix}{pmid:}
\providecommand{\doi}[1]{\href{http://dx.doi.org/#1}{\path{#1}}}
\providecommand{\Pubmed}[1]{\href{pmid:#1}{\path{#1}}}
\providecommand{\bibinfo}[2]{#2}
\ifx\xfnm\relax \def\xfnm[#1]{\unskip,\space#1}\fi
\bibitem[{{Alesina} et~al.(2019){Alesina}, {Cabot}, {Buchschacher} and
  {Burnier}}]{2019ASPC..523...25A}
\bibinfo{author}{{Alesina}, F.}, \bibinfo{author}{{Cabot}, F.},
  \bibinfo{author}{{Buchschacher}, N.}, \bibinfo{author}{{Burnier}, J.},
  \bibinfo{year}{2019}.
\newblock \bibinfo{title}{{Exoplanets Data Visualization in Multi-dimensional
  Plots using Virtual Reality in DACE}}, in: \bibinfo{editor}{{Teuben}, P.J.},
  \bibinfo{editor}{{Pound}, M.W.}, \bibinfo{editor}{{Thomas}, B.A.},
  \bibinfo{editor}{{Warner}, E.M.} (Eds.), \bibinfo{booktitle}{Astronomical
  Data Analysis Software and Systems XXVII}, p.~\bibinfo{pages}{25}.
\bibitem[{{Alho} et~al.(2019){Alho}, {Kallio} and
  {J{\"a}rvinen}}]{2019EGUGA..2112701A}
\bibinfo{author}{{Alho}, M.}, \bibinfo{author}{{Kallio}, E.},
  \bibinfo{author}{{J{\"a}rvinen}, R.}, \bibinfo{year}{2019}.
\newblock \bibinfo{title}{{Virtual Planetarium - Space Physics in Virtual
  Reality}}, in: \bibinfo{booktitle}{EGU General Assembly Conference
  Abstracts}, p. \bibinfo{pages}{12701}.
\bibitem[{{Arcand} et~al.(2018){Arcand}, {Jiang}, {Price}, {Watzke}, {Sgouros}
  and {Edmonds}}]{2018CAPJ...24...17A}
\bibinfo{author}{{Arcand}, K.K.}, \bibinfo{author}{{Jiang}, E.},
  \bibinfo{author}{{Price}, S.}, \bibinfo{author}{{Watzke}, M.},
  \bibinfo{author}{{Sgouros}, T.}, \bibinfo{author}{{Edmonds}, P.},
  \bibinfo{year}{2018}.
\newblock \bibinfo{title}{{Walking Through an Exploded Star: Rendering
  Supernova Remnant Cassiopeia A into Virtual Reality}}.
\newblock \bibinfo{journal}{Communicating Astronomy with the Public Journal}
  \bibinfo{volume}{24}, \bibinfo{pages}{17}.
\bibitem[{{Astropy Collaboration} et~al.(2013){Astropy Collaboration},
  {Robitaille}, {Tollerud}, {Greenfield}, {Droettboom}, {Bray}, {Aldcroft},
  {Davis}, {Ginsburg}, {Price-Whelan}, {Kerzendorf}, {Conley}, {Crighton},
  {Barbary}, {Muna}, {Ferguson}, {Grollier}, {Parikh}, {Nair}, {Unther},
  {Deil}, {Woillez}, {Conseil}, {Kramer}, {Turner}, {Singer}, {Fox}, {Weaver},
  {Zabalza}, {Edwards}, {Azalee Bostroem}, {Burke}, {Casey}, {Crawford},
  {Dencheva}, {Ely}, {Jenness}, {Labrie}, {Lim}, {Pierfederici}, {Pontzen},
  {Ptak}, {Refsdal}, {Servillat} and {Streicher}}]{2013A&A...558A..33A}
\bibinfo{author}{{Astropy Collaboration}}, \bibinfo{author}{{Robitaille},
  T.P.}, \bibinfo{author}{{Tollerud}, E.J.}, \bibinfo{author}{{Greenfield},
  P.}, \bibinfo{author}{{Droettboom}, M.}, \bibinfo{author}{{Bray}, E.},
  \bibinfo{author}{{Aldcroft}, T.}, \bibinfo{author}{{Davis}, M.},
  \bibinfo{author}{{Ginsburg}, A.}, \bibinfo{author}{{Price-Whelan}, A.M.},
  \bibinfo{author}{{Kerzendorf}, W.E.}, \bibinfo{author}{{Conley}, A.},
  \bibinfo{author}{{Crighton}, N.}, \bibinfo{author}{{Barbary}, K.},
  \bibinfo{author}{{Muna}, D.}, \bibinfo{author}{{Ferguson}, H.},
  \bibinfo{author}{{Grollier}, F.}, \bibinfo{author}{{Parikh}, M.M.},
  \bibinfo{author}{{Nair}, P.H.}, \bibinfo{author}{{Unther}, H.M.},
  \bibinfo{author}{{Deil}, C.}, \bibinfo{author}{{Woillez}, J.},
  \bibinfo{author}{{Conseil}, S.}, \bibinfo{author}{{Kramer}, R.},
  \bibinfo{author}{{Turner}, J.E.H.}, \bibinfo{author}{{Singer}, L.},
  \bibinfo{author}{{Fox}, R.}, \bibinfo{author}{{Weaver}, B.A.},
  \bibinfo{author}{{Zabalza}, V.}, \bibinfo{author}{{Edwards}, Z.I.},
  \bibinfo{author}{{Azalee Bostroem}, K.}, \bibinfo{author}{{Burke}, D.J.},
  \bibinfo{author}{{Casey}, A.R.}, \bibinfo{author}{{Crawford}, S.M.},
  \bibinfo{author}{{Dencheva}, N.}, \bibinfo{author}{{Ely}, J.},
  \bibinfo{author}{{Jenness}, T.}, \bibinfo{author}{{Labrie}, K.},
  \bibinfo{author}{{Lim}, P.L.}, \bibinfo{author}{{Pierfederici}, F.},
  \bibinfo{author}{{Pontzen}, A.}, \bibinfo{author}{{Ptak}, A.},
  \bibinfo{author}{{Refsdal}, B.}, \bibinfo{author}{{Servillat}, M.},
  \bibinfo{author}{{Streicher}, O.}, \bibinfo{year}{2013}.
\newblock \bibinfo{title}{{Astropy: A community Python package for astronomy}}.
\newblock \bibinfo{journal}{A\&A} \bibinfo{volume}{558}, \bibinfo{pages}{A33}.
\bibitem[{{Ball} and {Brunner}(2010)}]{Ball2010}
\bibinfo{author}{{Ball}, N.M.}, \bibinfo{author}{{Brunner}, R.J.},
  \bibinfo{year}{2010}.
\newblock \bibinfo{title}{{Data Mining and Machine Learning in Astronomy}}.
\newblock \bibinfo{journal}{International Journal of Modern Physics D}
  \bibinfo{volume}{19}, \bibinfo{pages}{1049--1106}.
\bibitem[{{Baracaglia} and {Vogt}(2020)}]{Baracaglia20}
\bibinfo{author}{{Baracaglia}, E.}, \bibinfo{author}{{Vogt}, F.P.A.},
  \bibinfo{year}{2020}.
\newblock \bibinfo{title}{{E0102-VR: Exploring the scientific potential of
  Virtual Reality for observational astrophysics}}.
\newblock \bibinfo{journal}{Astronomy and Computing} \bibinfo{volume}{30},
  \bibinfo{pages}{100352}.
\bibitem[{{Barnes} et~al.(2006){Barnes}, {Fluke}, {Bourke} and
  {Parry}}]{Barnes06}
\bibinfo{author}{{Barnes}, D.G.}, \bibinfo{author}{{Fluke}, C.J.},
  \bibinfo{author}{{Bourke}, P.D.}, \bibinfo{author}{{Parry}, O.T.},
  \bibinfo{year}{2006}.
\newblock \bibinfo{title}{{An Advanced, Three-Dimensional Plotting Library for
  Astronomy}}.
\newblock \bibinfo{journal}{PASA} \bibinfo{volume}{23},
  \bibinfo{pages}{82--93}.
\bibitem[{{Baron}(2019)}]{Baron19}
\bibinfo{author}{{Baron}, D.}, \bibinfo{year}{2019}.
\newblock \bibinfo{title}{{Machine Learning in Astronomy: a practical
  overview}}.
\newblock \bibinfo{journal}{arXiv e-prints}
  \href{http://arxiv.org/abs/1904.07248}{\tt arXiv:1904.07248}.
\bibitem[{{Bertin} and {Arnouts}(1996)}]{Bertin96}
\bibinfo{author}{{Bertin}, E.}, \bibinfo{author}{{Arnouts}, S.},
  \bibinfo{year}{1996}.
\newblock \bibinfo{title}{{SExtractor: Software for source extraction.}}
\newblock \bibinfo{journal}{Astronomy \& Astrophysics Supplement}
  \bibinfo{volume}{117}, \bibinfo{pages}{393--404}.
\bibitem[{{Bocchino} et~al.(2019){Bocchino}, {Orlando}, {Pillitteri}, {Miceli}
  and {Peres}}]{2019sros.confE..76B}
\bibinfo{author}{{Bocchino}, F.}, \bibinfo{author}{{Orlando}, S.},
  \bibinfo{author}{{Pillitteri}, I.}, \bibinfo{author}{{Miceli}, M.},
  \bibinfo{author}{{Peres}, G.}, \bibinfo{year}{2019}.
\newblock \bibinfo{title}{{A Virtual Reality Environment for Scientific
  Exploitation of 3D MHD Astrophysical Simulations}}, in:
  \bibinfo{booktitle}{Supernova Remnants: An Odyssey in Space after Stellar
  Death II}, p.~\bibinfo{pages}{76}.
\bibitem[{Brunner et~al.(2002)Brunner, Djorgovski, Prince and
  Szalay}]{Brunner2002}
\bibinfo{author}{Brunner, R.J.}, \bibinfo{author}{Djorgovski, S.G.},
  \bibinfo{author}{Prince, T.A.}, \bibinfo{author}{Szalay, A.S.},
  \bibinfo{year}{2002}.
\newblock \bibinfo{title}{Massive datasets in astronomy}, in:
  \bibinfo{editor}{Abello, J.}, \bibinfo{editor}{Pardalos, P.M.},
  \bibinfo{editor}{Resende, M.G.C.} (Eds.), \bibinfo{booktitle}{Handbook of
  Massive Data Sets}, \bibinfo{publisher}{Springer US},
  \bibinfo{address}{Boston, MA}. pp. \bibinfo{pages}{931--979}.
\bibitem[{Bryson(1996)}]{Bryson96}
\bibinfo{author}{Bryson, S.}, \bibinfo{year}{1996}.
\newblock \bibinfo{title}{Virtual reality in scientific visualization}.
\newblock \bibinfo{journal}{Commun. ACM} \bibinfo{volume}{39},
  \bibinfo{pages}{62–71}.
\bibitem[{{Caravaca} et~al.(2020){Caravaca}, {Le Mou{\'e}lic}, {Mangold},
  {L'Haridon}, {Le Deit} and {Mass{\'e}}}]{2020P-SS..18204808C}
\bibinfo{author}{{Caravaca}, G.}, \bibinfo{author}{{Le Mou{\'e}lic}, S.},
  \bibinfo{author}{{Mangold}, N.}, \bibinfo{author}{{L'Haridon}, J.},
  \bibinfo{author}{{Le Deit}, L.}, \bibinfo{author}{{Mass{\'e}}, M.},
  \bibinfo{year}{2020}.
\newblock \bibinfo{title}{{3D digital outcrop model reconstruction of the
  Kimberley outcrop (Gale crater, Mars) and its integration into Virtual
  Reality for simulated geological analysis}}.
\newblock \bibinfo{journal}{Planetary and Space Science} \bibinfo{volume}{182},
  \bibinfo{pages}{104808}.
\bibitem[{Castet et~al.(2024)Castet, Termoz-Masson, Vizcay, Delachambre,
  Myrodia, Aguilar, Matonti and Kornprobst}]{10.1167/jov.24.4.19}
\bibinfo{author}{Castet, E.}, \bibinfo{author}{Termoz-Masson, J.},
  \bibinfo{author}{Vizcay, S.}, \bibinfo{author}{Delachambre, J.},
  \bibinfo{author}{Myrodia, V.}, \bibinfo{author}{Aguilar, C.},
  \bibinfo{author}{Matonti, F.}, \bibinfo{author}{Kornprobst, P.},
  \bibinfo{year}{2024}.
\newblock \bibinfo{title}{{PTVR – A software in Python to make virtual
  reality experiments easier to build and more reproducible}}.
\newblock \bibinfo{journal}{Journal of Vision} \bibinfo{volume}{24},
  \bibinfo{pages}{19--19}.
\bibitem[{{Clampin} et~al.(2002){Clampin}, {Sirianni}, {Hartig}, {Ford},
  {Illingworth}, {Burmester}, {Koldewynd}, {Martel}, {Riess}, {Schrein} and
  {Sullivan}}]{Clampin02}
\bibinfo{author}{{Clampin}, M.}, \bibinfo{author}{{Sirianni}, M.},
  \bibinfo{author}{{Hartig}, G.F.}, \bibinfo{author}{{Ford}, H.C.},
  \bibinfo{author}{{Illingworth}, G.D.}, \bibinfo{author}{{Burmester}, B.},
  \bibinfo{author}{{Koldewynd}, W.}, \bibinfo{author}{{Martel}, A.R.},
  \bibinfo{author}{{Riess}, A.}, \bibinfo{author}{{Schrein}, R.J.},
  \bibinfo{author}{{Sullivan}, P.C.}, \bibinfo{year}{2002}.
\newblock \bibinfo{title}{{In-flight Performance of the Advanced Camera for
  Surveys CCDs}}.
\newblock \bibinfo{journal}{Experimental Astronomy} \bibinfo{volume}{14},
  \bibinfo{pages}{107--115}.
\bibitem[{{Cooke} et~al.(2019){Cooke}, {D{\'\i}az-Merced}, {Foran}, {Hannam}
  and {Garcia}}]{2019IAUS..339..251C}
\bibinfo{author}{{Cooke}, J.}, \bibinfo{author}{{D{\'\i}az-Merced}, W.},
  \bibinfo{author}{{Foran}, G.}, \bibinfo{author}{{Hannam}, J.},
  \bibinfo{author}{{Garcia}, B.}, \bibinfo{year}{2019}.
\newblock \bibinfo{title}{{Exploring Data Sonification to Enable, Enhance, and
  Accelerate the Analysis of Big, Noisy, and Multi-Dimensional Data}}, in:
  \bibinfo{editor}{{Griffin}, R.E.} (Ed.), \bibinfo{booktitle}{Southern
  Horizons in Time-Domain Astronomy}, pp. \bibinfo{pages}{251--256}.
\bibitem[{Cruz-Neira et~al.(1992)Cruz-Neira, Sandin, DeFanti, Kenyon and
  Hart}]{Cruz92}
\bibinfo{author}{Cruz-Neira, C.}, \bibinfo{author}{Sandin, D.J.},
  \bibinfo{author}{DeFanti, T.A.}, \bibinfo{author}{Kenyon, R.V.},
  \bibinfo{author}{Hart, J.C.}, \bibinfo{year}{1992}.
\newblock \bibinfo{title}{The cave: Audio visual experience automatic virtual
  environment}.
\newblock \bibinfo{journal}{Commun. ACM} \bibinfo{volume}{35},
  \bibinfo{pages}{64–72}.
\bibitem[{{Davelaar} et~al.(2018){Davelaar}, {Bronzwaer}, {Kok}, {Younsi},
  {Mo{\'s}cibrodzka} and {Falcke}}]{2018ComAC...5....1D}
\bibinfo{author}{{Davelaar}, J.}, \bibinfo{author}{{Bronzwaer}, T.},
  \bibinfo{author}{{Kok}, D.}, \bibinfo{author}{{Younsi}, Z.},
  \bibinfo{author}{{Mo{\'s}cibrodzka}, M.}, \bibinfo{author}{{Falcke}, H.},
  \bibinfo{year}{2018}.
\newblock \bibinfo{title}{{Observing supermassive black holes in virtual
  reality}}.
\newblock \bibinfo{journal}{Computational Astrophysics and Cosmology}
  \bibinfo{volume}{5}, \bibinfo{pages}{1}.
\bibitem[{{de Koning} et~al.(2021){de Koning}, {Egiz}, {Kotecha}, {Ciuculete},
  {Ooi}, {Bankole}, {Erhabor}, {Higginbotham}, {Khan}, {Dalle}, {Sichimba},
  {Bandyopadhyay} and {Kanmounye}}]{10.3389}
\bibinfo{author}{{de Koning}, R.}, \bibinfo{author}{{Egiz}, A.},
  \bibinfo{author}{{Kotecha}, J.}, \bibinfo{author}{{Ciuculete}, A.C.},
  \bibinfo{author}{{Ooi}, S.Z.Y.}, \bibinfo{author}{{Bankole}, N.D.A.},
  \bibinfo{author}{{Erhabor}, J.}, \bibinfo{author}{{Higginbotham}, G.},
  \bibinfo{author}{{Khan}, M.}, \bibinfo{author}{{Dalle}, D.U.},
  \bibinfo{author}{{Sichimba}, D.}, \bibinfo{author}{{Bandyopadhyay}, S.},
  \bibinfo{author}{{Kanmounye}, U.S.}, \bibinfo{year}{2021}.
\newblock \bibinfo{title}{Survey fatigue during the covid-19 pandemic: An
  analysis of neurosurgery survey response rates}.
\newblock \bibinfo{journal}{Frontiers in Surgery} \bibinfo{volume}{8}.
\bibitem[{Deering(1992)}]{Deering92}
\bibinfo{author}{Deering, M.}, \bibinfo{year}{1992}.
\newblock \bibinfo{title}{High resolution virtual reality}.
\newblock \bibinfo{journal}{SIGGRAPH Comput. Graph.} \bibinfo{volume}{26},
  \bibinfo{pages}{195–202}.
\bibitem[{{Deg} et~al.(2023){Deg}, {Palleske}, {Spekkens}, {Wang}, {Jarrett},
  {English}, {Lin}, {Yeung}, {Mould}, {Catinella}, {D{\'e}nes}, {Elagali},
  {For}, {Kamphuis}, {Koribalski}, {Lee-Waddell}, {Murugeshan}, {Oh}, {Rhee},
  {Serra}, {Westmeier}, {Wong}, {Bekki}, {Bosma}, {Carignan}, {Holwerda} and
  {Yu}}]{Deg23}
\bibinfo{author}{{Deg}, N.}, \bibinfo{author}{{Palleske}, R.},
  \bibinfo{author}{{Spekkens}, K.}, \bibinfo{author}{{Wang}, J.},
  \bibinfo{author}{{Jarrett}, T.}, \bibinfo{author}{{English}, J.},
  \bibinfo{author}{{Lin}, X.}, \bibinfo{author}{{Yeung}, J.},
  \bibinfo{author}{{Mould}, J.R.}, \bibinfo{author}{{Catinella}, B.},
  \bibinfo{author}{{D{\'e}nes}, H.}, \bibinfo{author}{{Elagali}, A.},
  \bibinfo{author}{{For}, B.Q.}, \bibinfo{author}{{Kamphuis}, P.},
  \bibinfo{author}{{Koribalski}, B.S.}, \bibinfo{author}{{Lee-Waddell}, K.},
  \bibinfo{author}{{Murugeshan}, C.}, \bibinfo{author}{{Oh}, S.},
  \bibinfo{author}{{Rhee}, J.}, \bibinfo{author}{{Serra}, P.},
  \bibinfo{author}{{Westmeier}, T.}, \bibinfo{author}{{Wong}, O.I.},
  \bibinfo{author}{{Bekki}, K.}, \bibinfo{author}{{Bosma}, A.},
  \bibinfo{author}{{Carignan}, C.}, \bibinfo{author}{{Holwerda}, B.W.},
  \bibinfo{author}{{Yu}, N.}, \bibinfo{year}{2023}.
\newblock \bibinfo{title}{{WALLABY pilot survey: the potential polar ring
  galaxies NGC 4632 and NGC 6156}}.
\newblock \bibinfo{journal}{MNRAS} \bibinfo{volume}{525},
  \bibinfo{pages}{4663--4684}.
\bibitem[{{Djorgovski} et~al.(2022){Djorgovski}, {Mahabal}, {Graham},
  {Polsterer} and {Krone-Martins}}]{2022arXiv221201493D}
\bibinfo{author}{{Djorgovski}, S.G.}, \bibinfo{author}{{Mahabal}, A.A.},
  \bibinfo{author}{{Graham}, M.J.}, \bibinfo{author}{{Polsterer}, K.},
  \bibinfo{author}{{Krone-Martins}, A.}, \bibinfo{year}{2022}.
\newblock \bibinfo{title}{{Applications of AI in Astronomy}}.
\newblock \bibinfo{journal}{arXiv e-prints}
  \href{http://arxiv.org/abs/2212.01493}{\tt arXiv:2212.01493}.
\bibitem[{Donalek et~al.(2014)Donalek, Djorgovski, Cioc, Wang, Zhang, Lawler,
  Yeh, Mahabal, Graham, Drake, Davidoff, Norris and Longo}]{7004282}
\bibinfo{author}{Donalek, C.}, \bibinfo{author}{Djorgovski, S.G.},
  \bibinfo{author}{Cioc, A.}, \bibinfo{author}{Wang, A.},
  \bibinfo{author}{Zhang, J.}, \bibinfo{author}{Lawler, E.},
  \bibinfo{author}{Yeh, S.}, \bibinfo{author}{Mahabal, A.},
  \bibinfo{author}{Graham, M.}, \bibinfo{author}{Drake, A.},
  \bibinfo{author}{Davidoff, S.}, \bibinfo{author}{Norris, J.S.},
  \bibinfo{author}{Longo, G.}, \bibinfo{year}{2014}.
\newblock \bibinfo{title}{Immersive and collaborative data visualization using
  virtual reality platforms}, in: \bibinfo{booktitle}{2014 IEEE International
  Conference on Big Data (Big Data)}, pp. \bibinfo{pages}{609--614}.
\bibitem[{{Elgner} et~al.(2017){Elgner}, {Adeli}, {Gwinner}, {Preusker},
  {Kersten}, {Matz}, {Roatsch}, {Jaumann} and {Oberst}}]{2017EPSC...11..481E}
\bibinfo{author}{{Elgner}, S.}, \bibinfo{author}{{Adeli}, S.},
  \bibinfo{author}{{Gwinner}, K.}, \bibinfo{author}{{Preusker}, F.},
  \bibinfo{author}{{Kersten}, E.}, \bibinfo{author}{{Matz}, K.D.},
  \bibinfo{author}{{Roatsch}, T.}, \bibinfo{author}{{Jaumann}, R.},
  \bibinfo{author}{{Oberst}, J.}, \bibinfo{year}{2017}.
\newblock \bibinfo{title}{{Visualizing planetary data by using 3D engines}},
  in: \bibinfo{booktitle}{European Planetary Science Congress}, pp.
  \bibinfo{pages}{EPSC2017--481}.
\bibitem[{{Febretti} et~al.(2013){Febretti}, {Nishimoto}, {Thigpen},
  {Talandis}, {Long}, {Pirtle}, {Peterka}, {Verlo}, {Brown}, {Plepys},
  {Sandin}, {Renambot}, {Johnson} and {Leigh}}]{Febretti13}
\bibinfo{author}{{Febretti}, A.}, \bibinfo{author}{{Nishimoto}, A.},
  \bibinfo{author}{{Thigpen}, T.}, \bibinfo{author}{{Talandis}, J.},
  \bibinfo{author}{{Long}, L.}, \bibinfo{author}{{Pirtle}, J.D.},
  \bibinfo{author}{{Peterka}, T.}, \bibinfo{author}{{Verlo}, A.},
  \bibinfo{author}{{Brown}, M.}, \bibinfo{author}{{Plepys}, D.},
  \bibinfo{author}{{Sandin}, D.}, \bibinfo{author}{{Renambot}, L.},
  \bibinfo{author}{{Johnson}, A.}, \bibinfo{author}{{Leigh}, J.},
  \bibinfo{year}{2013}.
\newblock \bibinfo{title}{{CAVE2: a hybrid reality environment for immersive
  simulation and information analysis}}, in: \bibinfo{editor}{{Dolinsky}, M.},
  \bibinfo{editor}{{McDowall}, I.E.} (Eds.), \bibinfo{booktitle}{The
  Engineering Reality of Virtual Reality 2013}, p. \bibinfo{pages}{864903}.
\bibitem[{{Ferrand} et~al.(2016){Ferrand}, {English} and {Irani}}]{Ferrand16}
\bibinfo{author}{{Ferrand}, G.}, \bibinfo{author}{{English}, J.},
  \bibinfo{author}{{Irani}, P.}, \bibinfo{year}{2016}.
\newblock \bibinfo{title}{{3D visualization of astronomy data cubes using
  immersive displays}}.
\newblock \bibinfo{journal}{arXiv e-prints}
  \href{http://arxiv.org/abs/1607.08874}{\tt arXiv:1607.08874}.
\bibitem[{{Ferrand} and {Warren}(2018)}]{2018CAPJ...24...25F}
\bibinfo{author}{{Ferrand}, G.}, \bibinfo{author}{{Warren}, D.},
  \bibinfo{year}{2018}.
\newblock \bibinfo{title}{{Engaging the Public with Supernova and Supernova
  Remnant Research Using Virtual Reality}}.
\newblock \bibinfo{journal}{Communicating Astronomy with the Public Journal}
  \bibinfo{volume}{24}, \bibinfo{pages}{25}.
\bibitem[{{Flaugher}(2006)}]{Flaugher06}
\bibinfo{author}{{Flaugher}, B.}, \bibinfo{year}{2006}.
\newblock \bibinfo{title}{{The Dark Energy Survey instrument design}}, in:
  \bibinfo{editor}{{McLean}, I.S.}, \bibinfo{editor}{{Iye}, M.} (Eds.),
  \bibinfo{booktitle}{Society of Photo-Optical Instrumentation Engineers (SPIE)
  Conference Series}, p. \bibinfo{pages}{62692C}.
\bibitem[{{Fluke} and {Barnes}(2018)}]{Fluke18}
\bibinfo{author}{{Fluke}, C.J.}, \bibinfo{author}{{Barnes}, D.G.},
  \bibinfo{year}{2018}.
\newblock \bibinfo{title}{{Immersive Virtual Reality Experiences for All-Sky
  Data}}.
\newblock \bibinfo{journal}{PASA} \bibinfo{volume}{35}, \bibinfo{pages}{e026}.
\bibitem[{{Fluke} et~al.(2006){Fluke}, {Bourke} and {O'Donovan}}]{Fluke06}
\bibinfo{author}{{Fluke}, C.J.}, \bibinfo{author}{{Bourke}, P.D.},
  \bibinfo{author}{{O'Donovan}, D.}, \bibinfo{year}{2006}.
\newblock \bibinfo{title}{{Future Directions in Astronomy Visualization}}.
\newblock \bibinfo{journal}{PASA} \bibinfo{volume}{23},
  \bibinfo{pages}{12--24}.
\bibitem[{{Fluke} and {Jacobs}(2020)}]{Fluke20a}
\bibinfo{author}{{Fluke}, C.J.}, \bibinfo{author}{{Jacobs}, C.},
  \bibinfo{year}{2020}.
\newblock \bibinfo{title}{{Surveying the reach and maturity of machine learning
  and artificial intelligence in astronomy}}.
\newblock \bibinfo{journal}{WIREs Data Mining and Knowledge Discovery}
  \bibinfo{volume}{10}, \bibinfo{pages}{e1349}.
\bibitem[{{Fluke} et~al.(2023){Fluke}, {Vohl}, {Kilborn} and
  {Murugeshan}}]{Fluke2023}
\bibinfo{author}{{Fluke}, C.J.}, \bibinfo{author}{{Vohl}, D.},
  \bibinfo{author}{{Kilborn}, V.A.}, \bibinfo{author}{{Murugeshan}, C.},
  \bibinfo{year}{2023}.
\newblock \bibinfo{title}{{Survey-scale discovery-based research processes:
  Evaluating a bespoke visualisation environment for astronomical survey
  data}}.
\newblock \bibinfo{journal}{PASA} \bibinfo{volume}{40}, \bibinfo{pages}{e035}.
\bibitem[{{Fomalont}(1982)}]{Fomalont82}
\bibinfo{author}{{Fomalont}, E.}, \bibinfo{year}{1982}.
\newblock \bibinfo{title}{{Image Display and Analysis}}, in:
  \bibinfo{booktitle}{Synthesis Mapping}, p.~\bibinfo{pages}{11}.
\bibitem[{{Glowacki} et~al.(2024){Glowacki}, {Albrow}, {Reynolds}, {Elson},
  {Mahony} and {Allison}}]{Glowacki24}
\bibinfo{author}{{Glowacki}, M.}, \bibinfo{author}{{Albrow}, L.},
  \bibinfo{author}{{Reynolds}, T.}, \bibinfo{author}{{Elson}, E.},
  \bibinfo{author}{{Mahony}, E.K.}, \bibinfo{author}{{Allison}, J.R.},
  \bibinfo{year}{2024}.
\newblock \bibinfo{title}{{A serendipitous discovery of H I-rich galaxy groups
  with MeerKAT}}.
\newblock \bibinfo{journal}{MNRAS} \bibinfo{volume}{529},
  \bibinfo{pages}{3469--3483}.
\bibitem[{{Hassan} and {Fluke}(2011)}]{Hassan11}
\bibinfo{author}{{Hassan}, A.}, \bibinfo{author}{{Fluke}, C.J.},
  \bibinfo{year}{2011}.
\newblock \bibinfo{title}{{Scientific Visualization in Astronomy: Towards the
  Petascale Astronomy Era}}.
\newblock \bibinfo{journal}{PASA} \bibinfo{volume}{28},
  \bibinfo{pages}{150--170}.
\bibitem[{{Higashio} et~al.(2022){Higashio}, {Kuchner}, {Silverberg}, {Brandt},
  {Grubb}, {Gagn{\'e}}, {Debes}, {Schlieder}, {Wisniewski}, {Slocum}, {Bans},
  {Bhattacharjee}, {Biggs}, {Bosch}, {Cernohous}, {Doll}, {Durantini Luca},
  {Enachioaie}, {Griffith}, {Hamilton}, {Holden}, {Hyogo}, {Jung}, {Lau},
  {Pi{\~n}eiro}, {Piipuu}, {Stiller} and {Disk Detective
  Collaboration}}]{Higashio22}
\bibinfo{author}{{Higashio}, S.}, \bibinfo{author}{{Kuchner}, M.J.},
  \bibinfo{author}{{Silverberg}, S.M.}, \bibinfo{author}{{Brandt}, M.A.},
  \bibinfo{author}{{Grubb}, T.G.}, \bibinfo{author}{{Gagn{\'e}}, J.},
  \bibinfo{author}{{Debes}, J.H.}, \bibinfo{author}{{Schlieder}, J.},
  \bibinfo{author}{{Wisniewski}, J.P.}, \bibinfo{author}{{Slocum}, S.},
  \bibinfo{author}{{Bans}, A.S.}, \bibinfo{author}{{Bhattacharjee}, S.},
  \bibinfo{author}{{Biggs}, J.R.}, \bibinfo{author}{{Bosch}, M.K.D.},
  \bibinfo{author}{{Cernohous}, T.}, \bibinfo{author}{{Doll}, K.},
  \bibinfo{author}{{Durantini Luca}, H.A.}, \bibinfo{author}{{Enachioaie}, A.},
  \bibinfo{author}{{Griffith}, P.}, \bibinfo{author}{{Hamilton}, J.},
  \bibinfo{author}{{Holden}, J.}, \bibinfo{author}{{Hyogo}, M.},
  \bibinfo{author}{{Jung}, D.}, \bibinfo{author}{{Lau}, L.},
  \bibinfo{author}{{Pi{\~n}eiro}, F.}, \bibinfo{author}{{Piipuu}, A.},
  \bibinfo{author}{{Stiller}, L.}, \bibinfo{author}{{Disk Detective
  Collaboration}}, \bibinfo{year}{2022}.
\newblock \bibinfo{title}{{Disks in Nearby Young Stellar Associations Found Via
  Virtual Reality}}.
\newblock \bibinfo{journal}{ApJS} \bibinfo{volume}{933}, \bibinfo{pages}{13}.
\bibitem[{Holliman et~al.(2011)Holliman, Dodgson, Favalora and
  Pockett}]{Holliman11}
\bibinfo{author}{Holliman, N.S.}, \bibinfo{author}{Dodgson, N.A.},
  \bibinfo{author}{Favalora, G.E.}, \bibinfo{author}{Pockett, L.},
  \bibinfo{year}{2011}.
\newblock \bibinfo{title}{Three-dimensional displays: A review and applications
  analysis}.
\newblock \bibinfo{journal}{IEEE Transactions on Broadcasting}
  \bibinfo{volume}{57}, \bibinfo{pages}{362--371}.
\bibitem[{{Impey} and {Danehy}(2022)}]{2022CAPJ...31...28I}
\bibinfo{author}{{Impey}, C.}, \bibinfo{author}{{Danehy}, A.},
  \bibinfo{year}{2022}.
\newblock \bibinfo{title}{{Exploring the Frontiers of Space in 3D: Immersive
  Virtual Reality for Astronomy Outreach}}.
\newblock \bibinfo{journal}{Communicating Astronomy with the Public Journal}
  \bibinfo{volume}{31}, \bibinfo{pages}{28}.
\bibitem[{{Jarrett} et~al.(2021){Jarrett}, {Comrie}, {Marchetti}, {Sivitilli},
  {Macfarlane}, {Vitello}, {Becciani}, {Taylor}, {van der Hulst}, {Serra},
  {Katz} and {Cluver}}]{Jarrett21}
\bibinfo{author}{{Jarrett}, T.H.}, \bibinfo{author}{{Comrie}, A.},
  \bibinfo{author}{{Marchetti}, L.}, \bibinfo{author}{{Sivitilli}, A.},
  \bibinfo{author}{{Macfarlane}, S.}, \bibinfo{author}{{Vitello}, F.},
  \bibinfo{author}{{Becciani}, U.}, \bibinfo{author}{{Taylor}, A.R.},
  \bibinfo{author}{{van der Hulst}, J.M.}, \bibinfo{author}{{Serra}, P.},
  \bibinfo{author}{{Katz}, N.}, \bibinfo{author}{{Cluver}, M.E.},
  \bibinfo{year}{2021}.
\newblock \bibinfo{title}{{Exploring and interrogating astrophysical data in
  virtual reality}}.
\newblock \bibinfo{journal}{Astronomy and Computing} \bibinfo{volume}{37},
  \bibinfo{pages}{100502}.
\bibitem[{{Jonas} and {MeerKAT Team}(2016)}]{Jonas2016}
\bibinfo{author}{{Jonas}, J.}, \bibinfo{author}{{MeerKAT Team}},
  \bibinfo{year}{2016}.
\newblock \bibinfo{title}{{The MeerKAT Radio Telescope}}, in:
  \bibinfo{booktitle}{MeerKAT Science: On the Pathway to the SKA},
  p.~\bibinfo{pages}{1}.
\bibitem[{{Kaluza} et~al.(2019){Kaluza}, {Moresi}, {Mansour} and
  {Barnes}}]{2019zndo...3348451K}
\bibinfo{author}{{Kaluza}, O.}, \bibinfo{author}{{Moresi}, L.},
  \bibinfo{author}{{Mansour}, J.}, \bibinfo{author}{{Barnes}, D.G.},
  \bibinfo{year}{2019}.
\newblock \bibinfo{title}{{OKaluza/LavaVu: v1.4.3}}.
\bibitem[{Kersting et~al.(2024)Kersting, Bondell, Steir and Myers}]{Kersting24}
\bibinfo{author}{Kersting, M.}, \bibinfo{author}{Bondell, J.},
  \bibinfo{author}{Steir, R.}, \bibinfo{author}{Myers, M.},
  \bibinfo{year}{2024}.
\newblock \bibinfo{title}{Virtual reality in astronomy education: reflecting on
  design principles through a dialogue between researchers and practitioners}.
\newblock \bibinfo{journal}{International Journal of Science Education, Part B}
  \bibinfo{volume}{14}, \bibinfo{pages}{157--176}.
\bibitem[{{Kleiner} et~al.(2021){Kleiner}, {Serra}, {Maccagni}, {Venhola},
  {Morokuma-Matsui}, {Peletier}, {Iodice}, {Raj}, {de Blok}, {Comrie},
  {J{\'o}zsa}, {Kamphuis}, {Loni}, {Loubser}, {Moln{\'a}r}, {Passmoor},
  {Ramatsoku}, {Sivitilli}, {Smirnov}, {Thorat} and {Vitello}}]{Kleiner21}
\bibinfo{author}{{Kleiner}, D.}, \bibinfo{author}{{Serra}, P.},
  \bibinfo{author}{{Maccagni}, F.M.}, \bibinfo{author}{{Venhola}, A.},
  \bibinfo{author}{{Morokuma-Matsui}, K.}, \bibinfo{author}{{Peletier}, R.},
  \bibinfo{author}{{Iodice}, E.}, \bibinfo{author}{{Raj}, M.A.},
  \bibinfo{author}{{de Blok}, W.J.G.}, \bibinfo{author}{{Comrie}, A.},
  \bibinfo{author}{{J{\'o}zsa}, G.I.G.}, \bibinfo{author}{{Kamphuis}, P.},
  \bibinfo{author}{{Loni}, A.}, \bibinfo{author}{{Loubser}, S.I.},
  \bibinfo{author}{{Moln{\'a}r}, D.C.}, \bibinfo{author}{{Passmoor}, S.S.},
  \bibinfo{author}{{Ramatsoku}, M.}, \bibinfo{author}{{Sivitilli}, A.},
  \bibinfo{author}{{Smirnov}, O.}, \bibinfo{author}{{Thorat}, K.},
  \bibinfo{author}{{Vitello}, F.}, \bibinfo{year}{2021}.
\newblock \bibinfo{title}{{A MeerKAT view of pre-processing in the Fornax A
  group}}.
\newblock \bibinfo{journal}{A\&A} \bibinfo{volume}{648}, \bibinfo{pages}{A32}.
\bibitem[{{Krieger} et~al.(2023){Krieger}, {LeBlanc}, {Waterman}, {Reisner},
  {Testa} and {Chen}}]{Krieger23}
\bibinfo{author}{{Krieger}, N.}, \bibinfo{author}{{LeBlanc}, M.},
  \bibinfo{author}{{Waterman}, P.}, \bibinfo{author}{{Reisner}, S.},
  \bibinfo{author}{{Testa}, C.}, \bibinfo{author}{{Chen}, J.},
  \bibinfo{year}{2023}.
\newblock \bibinfo{title}{{Decreasing Survey Response Rates in the Time of
  COVID-19: Implications for Analyses of Population Health and Health
  Inequities}}.
\newblock \bibinfo{journal}{American Journal of Public Health}
  \bibinfo{volume}{113}, \bibinfo{pages}{667--670}.
\bibitem[{{Kuchner} et~al.(2020){Kuchner}, {Higashio} and
  {Brandt}}]{2020AAS...23535704K}
\bibinfo{author}{{Kuchner}, M.J.}, \bibinfo{author}{{Higashio}, S.},
  \bibinfo{author}{{Brandt}, M.A.}, \bibinfo{year}{2020}.
\newblock \bibinfo{title}{{Disks-Hosting Members of Columba-Carina Found Using
  Disk Detective and Virtual Reality}}, in: \bibinfo{booktitle}{American
  Astronomical Society Meeting Abstracts}, p. \bibinfo{pages}{357.04}.
\bibitem[{{Lambert} et~al.(2020){Lambert}, {Kraan-Korteweg}, {Jarrett} and
  {Macri}}]{2020MNRAS.497.2954L}
\bibinfo{author}{{Lambert}, T.S.}, \bibinfo{author}{{Kraan-Korteweg}, R.C.},
  \bibinfo{author}{{Jarrett}, T.H.}, \bibinfo{author}{{Macri}, L.M.},
  \bibinfo{year}{2020}.
\newblock \bibinfo{title}{{The 2MASS redshift survey galaxy group catalogue
  derived from a graph-theory based friends-of-friends algorithm}}.
\newblock \bibinfo{journal}{MNRAS} \bibinfo{volume}{497},
  \bibinfo{pages}{2954--2973}.
\bibitem[{Lan et~al.(2021)Lan, Young, Anderson, Ynnerman, Bock, Borkin, Forbes,
  Kollmeier and Wang}]{Lan21}
\bibinfo{author}{Lan, F.}, \bibinfo{author}{Young, M.},
  \bibinfo{author}{Anderson, L.}, \bibinfo{author}{Ynnerman, A.},
  \bibinfo{author}{Bock, A.}, \bibinfo{author}{Borkin, M.A.},
  \bibinfo{author}{Forbes, A.G.}, \bibinfo{author}{Kollmeier, J.A.},
  \bibinfo{author}{Wang, B.}, \bibinfo{year}{2021}.
\newblock \bibinfo{title}{Visualization in astrophysics: Developing new
  methods, discovering our universe, and educating the earth}.
\newblock \bibinfo{journal}{Computer Graphics Forum} \bibinfo{volume}{40},
  \bibinfo{pages}{635--663}.
\bibitem[{{Le Mou{\'e}lic} et~al.(2019){Le Mou{\'e}lic}, {Caravaca}, {Mangold},
  {L'Haridon}, {Le Deit} and {Mass{\'e}}}]{2019EGUGA..21.3939L}
\bibinfo{author}{{Le Mou{\'e}lic}, S.}, \bibinfo{author}{{Caravaca}, G.},
  \bibinfo{author}{{Mangold}, N.}, \bibinfo{author}{{L'Haridon}, J.},
  \bibinfo{author}{{Le Deit}, L.}, \bibinfo{author}{{Mass{\'e}}, M.},
  \bibinfo{year}{2019}.
\newblock \bibinfo{title}{{Geologic mapping and stratigraphy of remote Martian
  outcrops using digital outcrop model and virtual reality: example of the
  Kimberley outcrop (Gale Crater, Mars)}}, in: \bibinfo{booktitle}{EGU General
  Assembly Conference Abstracts}, p. \bibinfo{pages}{3939}.
\bibitem[{{Le Mou{\'e}lic} et~al.(2018){Le Mou{\'e}lic}, {L'Haridon}, {Civet},
  {Mangold}, {Triantafyllou}, {Mass{\'e}}, {Le Menn} and
  {Beaunay}}]{2018EGUGA..2013366L}
\bibinfo{author}{{Le Mou{\'e}lic}, S.}, \bibinfo{author}{{L'Haridon}, J.},
  \bibinfo{author}{{Civet}, F.}, \bibinfo{author}{{Mangold}, N.},
  \bibinfo{author}{{Triantafyllou}, A.}, \bibinfo{author}{{Mass{\'e}}, M.},
  \bibinfo{author}{{Le Menn}, E.}, \bibinfo{author}{{Beaunay}, S.},
  \bibinfo{year}{2018}.
\newblock \bibinfo{title}{{Using virtual reality to investigate geological
  outcrops on planetary surfaces}}, in: \bibinfo{booktitle}{EGU General
  Assembly Conference Abstracts}, p. \bibinfo{pages}{13366}.
\bibitem[{{Lee} et~al.(2018){Lee}, {Krolewski}, {White}, {Schlegel}, {Nugent},
  {Hennawi}, {M{\"u}ller}, {Pan}, {Prochaska}, {Font-Ribera}, {Suzuki},
  {Glazebrook}, {Kacprzak}, {Kartaltepe}, {Koekemoer}, {Le F{\`e}vre},
  {Lemaux}, {Maier}, {Nanayakkara}, {Rich}, {Sanders}, {Salvato}, {Tasca} and
  {Tran}}]{2018ApJS..237...31L}
\bibinfo{author}{{Lee}, K.G.}, \bibinfo{author}{{Krolewski}, A.},
  \bibinfo{author}{{White}, M.}, \bibinfo{author}{{Schlegel}, D.},
  \bibinfo{author}{{Nugent}, P.E.}, \bibinfo{author}{{Hennawi}, J.F.},
  \bibinfo{author}{{M{\"u}ller}, T.}, \bibinfo{author}{{Pan}, R.},
  \bibinfo{author}{{Prochaska}, J.X.}, \bibinfo{author}{{Font-Ribera}, A.},
  \bibinfo{author}{{Suzuki}, N.}, \bibinfo{author}{{Glazebrook}, K.},
  \bibinfo{author}{{Kacprzak}, G.G.}, \bibinfo{author}{{Kartaltepe}, J.S.},
  \bibinfo{author}{{Koekemoer}, A.M.}, \bibinfo{author}{{Le F{\`e}vre}, O.},
  \bibinfo{author}{{Lemaux}, B.C.}, \bibinfo{author}{{Maier}, C.},
  \bibinfo{author}{{Nanayakkara}, T.}, \bibinfo{author}{{Rich}, R.M.},
  \bibinfo{author}{{Sanders}, D.B.}, \bibinfo{author}{{Salvato}, M.},
  \bibinfo{author}{{Tasca}, L.}, \bibinfo{author}{{Tran}, K.V.H.},
  \bibinfo{year}{2018}.
\newblock \bibinfo{title}{{First Data Release of the COSMOS
  Ly{\ensuremath{\alpha}} Mapping and Tomography Observations: 3D
  Ly{\ensuremath{\alpha}} Forest Tomography at $2.05 < z < 2.55$}}.
\newblock \bibinfo{journal}{ApJS} \bibinfo{volume}{237}, \bibinfo{pages}{31}.
\bibitem[{{Liu} et~al.(2021){Liu}, {Kerrison}, {Wurmser}, {Falcone},
  {Greenfield}, {Joyce}, {Marinelli}, {Ostling}, {Quayum}, {Williams}, {Liu}
  and {Y/dim Collaboration}}]{Liu21a}
\bibinfo{author}{{Liu}, A.G.}, \bibinfo{author}{{Kerrison}, N.},
  \bibinfo{author}{{Wurmser}, S.}, \bibinfo{author}{{Falcone}, J.},
  \bibinfo{author}{{Greenfield}, S.}, \bibinfo{author}{{Joyce}, M.},
  \bibinfo{author}{{Marinelli}, M.}, \bibinfo{author}{{Ostling}, W.},
  \bibinfo{author}{{Quayum}, R.}, \bibinfo{author}{{Williams}, R.},
  \bibinfo{author}{{Liu}, C.}, \bibinfo{author}{{Y/dim Collaboration}},
  \bibinfo{year}{2021}.
\newblock \bibinfo{title}{{Virtual Reality Mapping of E+A Galaxies and
  Candidates in the Coma Cluster and Other Nearby Rich Clusters of Galaxies}},
  in: \bibinfo{booktitle}{American Astronomical Society Meeting Abstracts}, p.
  \bibinfo{pages}{341.09}.
\bibitem[{{Liu} and {Liu}(2021)}]{Liu21b}
\bibinfo{author}{{Liu}, A.G.}, \bibinfo{author}{{Liu}, C.T.},
  \bibinfo{year}{2021}.
\newblock \bibinfo{title}{{Preliminary Results of Virtual Reality Mapping of
  E+A Galaxies and Candidates in Nearby Rich Clusters of Galaxies}}.
\newblock \bibinfo{journal}{Research Notes of the American Astronomical
  Society} \bibinfo{volume}{5}, \bibinfo{pages}{83}.
\bibitem[{{Longo} et~al.(2019){Longo}, {Mer{\'e}nyi} and
  {Ti{\v{n}}o}}]{Longo19}
\bibinfo{author}{{Longo}, G.}, \bibinfo{author}{{Mer{\'e}nyi}, E.},
  \bibinfo{author}{{Ti{\v{n}}o}, P.}, \bibinfo{year}{2019}.
\newblock \bibinfo{title}{{Foreword to the Focus Issue on Machine Intelligence
  in Astronomy and Astrophysics}}.
\newblock \bibinfo{journal}{PASP} \bibinfo{volume}{131},
  \bibinfo{pages}{100101}.
\bibitem[{{Maccagni} et~al.(2024){Maccagni}, {de Blok}, {Mancera Pi{\~n}a},
  {Ragusa}, {Iodice}, {Spavone}, {McGaugh}, {Oman}, {Oosterloo}, {Koribalski},
  {Kim}, {Adams}, {Amram}, {Bosma}, {Bigiel}, {Brinks}, {Chemin}, {Combes},
  {Gibson}, {Healy}, {Holwerda}, {J{\'o}zsa}, {Kamphuis}, {Kleiner},
  {Kurapati}, {Marasco}, {Spekkens}, {Veronese}, {Walter}, {Zabel} and
  {Zijlstra}}]{Maccagni24}
\bibinfo{author}{{Maccagni}, F.M.}, \bibinfo{author}{{de Blok}, W.J.G.},
  \bibinfo{author}{{Mancera Pi{\~n}a}, P.E.}, \bibinfo{author}{{Ragusa}, R.},
  \bibinfo{author}{{Iodice}, E.}, \bibinfo{author}{{Spavone}, M.},
  \bibinfo{author}{{McGaugh}, S.}, \bibinfo{author}{{Oman}, K.A.},
  \bibinfo{author}{{Oosterloo}, T.A.}, \bibinfo{author}{{Koribalski}, B.S.},
  \bibinfo{author}{{Kim}, M.}, \bibinfo{author}{{Adams}, E.A.K.},
  \bibinfo{author}{{Amram}, P.}, \bibinfo{author}{{Bosma}, A.},
  \bibinfo{author}{{Bigiel}, F.}, \bibinfo{author}{{Brinks}, E.},
  \bibinfo{author}{{Chemin}, L.}, \bibinfo{author}{{Combes}, F.},
  \bibinfo{author}{{Gibson}, B.}, \bibinfo{author}{{Healy}, J.},
  \bibinfo{author}{{Holwerda}, B.W.}, \bibinfo{author}{{J{\'o}zsa}, G.I.G.},
  \bibinfo{author}{{Kamphuis}, P.}, \bibinfo{author}{{Kleiner}, D.},
  \bibinfo{author}{{Kurapati}, S.}, \bibinfo{author}{{Marasco}, A.},
  \bibinfo{author}{{Spekkens}, K.}, \bibinfo{author}{{Veronese}, S.},
  \bibinfo{author}{{Walter}, F.}, \bibinfo{author}{{Zabel}, N.},
  \bibinfo{author}{{Zijlstra}, A.}, \bibinfo{year}{2024}.
\newblock \bibinfo{title}{{MHONGOOSE discovery of a gas-rich low surface
  brightness galaxy in the Dorado group}}.
\newblock \bibinfo{journal}{A\&A} \bibinfo{volume}{690}, \bibinfo{pages}{A69}.
\bibitem[{{Macri} et~al.(2019){Macri}, {Kraan-Korteweg}, {Lambert}, {Alonso},
  {Berlind}, {Calkins}, {Erdo{\u{g}}du}, {Falco}, {Jarrett} and
  {Mink}}]{2019ApJS..245....6M}
\bibinfo{author}{{Macri}, L.M.}, \bibinfo{author}{{Kraan-Korteweg}, R.C.},
  \bibinfo{author}{{Lambert}, T.}, \bibinfo{author}{{Alonso}, M.V.},
  \bibinfo{author}{{Berlind}, P.}, \bibinfo{author}{{Calkins}, M.},
  \bibinfo{author}{{Erdo{\u{g}}du}, P.}, \bibinfo{author}{{Falco}, E.E.},
  \bibinfo{author}{{Jarrett}, T.H.}, \bibinfo{author}{{Mink}, J.D.},
  \bibinfo{year}{2019}.
\newblock \bibinfo{title}{{The 2MASS Redshift Survey in the Zone of
  Avoidance}}.
\newblock \bibinfo{journal}{ApJS} \bibinfo{volume}{245}, \bibinfo{pages}{6}.
\bibitem[{{Makovoz} and {Marleau}(2005)}]{Makovoz05}
\bibinfo{author}{{Makovoz}, D.}, \bibinfo{author}{{Marleau}, F.R.},
  \bibinfo{year}{2005}.
\newblock \bibinfo{title}{{Point-Source Extraction with MOPEX}}.
\newblock \bibinfo{journal}{PASP} \bibinfo{volume}{117},
  \bibinfo{pages}{1113--1128}.
\bibitem[{{Marchetti} and {Jarrett}(2018)}]{Marchetti18}
\bibinfo{author}{{Marchetti}, L.}, \bibinfo{author}{{Jarrett}, T.H.},
  \bibinfo{year}{2018}.
\newblock \bibinfo{title}{{The Data2Dome Initiative at the Iziko Planetarium \&
  the IDIA Visualisation Lab}}, in: \bibinfo{booktitle}{BigSkyEarth Conference:
  AstroGeoInformatics}, p.~\bibinfo{pages}{7}.
\bibitem[{{Meade} et~al.(2014){Meade}, {Fluke}, {Manos} and
  {Sinnott}}]{Meade14}
\bibinfo{author}{{Meade}, B.F.}, \bibinfo{author}{{Fluke}, C.J.},
  \bibinfo{author}{{Manos}, S.}, \bibinfo{author}{{Sinnott}, R.O.},
  \bibinfo{year}{2014}.
\newblock \bibinfo{title}{{Are Tiled Display Walls Needed for Astronomy?}}
\newblock \bibinfo{journal}{PASA} \bibinfo{volume}{31}, \bibinfo{pages}{e033}.
\bibitem[{{Menon} et~al.(2020){Menon}, {Walker}, {Koris}, {Szafir} and
  {Burns}}]{2020LPICo2241.5017M}
\bibinfo{author}{{Menon}, M.S.}, \bibinfo{author}{{Walker}, M.E.},
  \bibinfo{author}{{Koris}, D.}, \bibinfo{author}{{Szafir}, D.},
  \bibinfo{author}{{Burns}, J.O.}, \bibinfo{year}{2020}.
\newblock \bibinfo{title}{{Virtual Reality Simulator for Telerobotics Research
  to Enable Artemis and the FARSIDE Low Frequency Radio Telescope}}.
\newblock \bibinfo{journal}{LPI Contributions} \bibinfo{volume}{2241},
  \bibinfo{pages}{5017}.
\bibitem[{{Milisavljevic} et~al.(2021){Milisavljevic}, {Sumner}, {Takahashi},
  {Martin}, {Drissen} and {Law}}]{Milisavljevic21}
\bibinfo{author}{{Milisavljevic}, D.}, \bibinfo{author}{{Sumner}, A.},
  \bibinfo{author}{{Takahashi}, G.}, \bibinfo{author}{{Martin}, T.},
  \bibinfo{author}{{Drissen}, L.}, \bibinfo{author}{{Law}, C.J.},
  \bibinfo{year}{2021}.
\newblock \bibinfo{title}{{Visualization and Collaborative Exploration of
  Complex Multi-dimensional Data Among Distant Individuals using Virtual
  Reality}}, in: \bibinfo{booktitle}{American Astronomical Society Meeting
  Abstracts}, p. \bibinfo{pages}{541.11}.
\bibitem[{{Mohan} and {Rafferty}(2015)}]{Mohan15}
\bibinfo{author}{{Mohan}, N.}, \bibinfo{author}{{Rafferty}, D.},
  \bibinfo{year}{2015}.
\newblock \bibinfo{title}{{PyBDSF: Python Blob Detection and Source Finder}}.
\newblock \bibinfo{howpublished}{Astrophysics Source Code Library, record
  ascl:1502.007}.
\bibitem[{{Moriwaki} et~al.(2023){Moriwaki}, {Nishimichi} and
  {Yoshida}}]{2023RPPh...86g6901M}
\bibinfo{author}{{Moriwaki}, K.}, \bibinfo{author}{{Nishimichi}, T.},
  \bibinfo{author}{{Yoshida}, N.}, \bibinfo{year}{2023}.
\newblock \bibinfo{title}{{Machine learning for observational cosmology}}.
\newblock \bibinfo{journal}{Reports on Progress in Physics}
  \bibinfo{volume}{86}, \bibinfo{pages}{076901}.
\bibitem[{{Moss} et~al.(2021){Moss}, {Adcock}, {Hotan}, {Kobayashi}, {Rees},
  {Si{\'e}gel}, {Tremblay} and {Trenham}}]{2021NatAs...5..213M}
\bibinfo{author}{{Moss}, V.A.}, \bibinfo{author}{{Adcock}, M.},
  \bibinfo{author}{{Hotan}, A.W.}, \bibinfo{author}{{Kobayashi}, R.},
  \bibinfo{author}{{Rees}, G.A.}, \bibinfo{author}{{Si{\'e}gel}, C.},
  \bibinfo{author}{{Tremblay}, C.D.}, \bibinfo{author}{{Trenham}, C.E.},
  \bibinfo{year}{2021}.
\newblock \bibinfo{title}{{Forging a path to a better normal for conferences
  and collaboration}}.
\newblock \bibinfo{journal}{Nature Astronomy} \bibinfo{volume}{5},
  \bibinfo{pages}{213--216}.
\bibitem[{{Moss} et~al.(2023){Moss}, {Rees}, {Hotan}, {Tasker}, {Kobayashi},
  {Kerrison}, {Amos} and {Ekers}}]{2023NatAs...7.1412M}
\bibinfo{author}{{Moss}, V.A.}, \bibinfo{author}{{Rees}, G.A.},
  \bibinfo{author}{{Hotan}, A.W.}, \bibinfo{author}{{Tasker}, E.J.},
  \bibinfo{author}{{Kobayashi}, R.}, \bibinfo{author}{{Kerrison}, E.F.},
  \bibinfo{author}{{Amos}, K.V.H.}, \bibinfo{author}{{Ekers}, R.D.},
  \bibinfo{year}{2023}.
\newblock \bibinfo{title}{{Going beyond being there to bring astronomy to the
  world}}.
\newblock \bibinfo{journal}{Nature Astronomy} \bibinfo{volume}{7},
  \bibinfo{pages}{1412--1414}.
\bibitem[{{Norris}(1994)}]{Norris94}
\bibinfo{author}{{Norris}, R.P.}, \bibinfo{year}{1994}.
\newblock \bibinfo{title}{{The Challenge of Astronomical Visualisation}}, in:
  \bibinfo{editor}{{Crabtree}, D.R.}, \bibinfo{editor}{{Hanisch}, R.J.},
  \bibinfo{editor}{{Barnes}, J.} (Eds.), \bibinfo{booktitle}{Astronomical Data
  Analysis Software and Systems III}, p.~\bibinfo{pages}{51}.
\bibitem[{{Orlando} et~al.(2019){Orlando}, {Pillitteri}, {Bocchino},
  {Daricello} and {Leonardi}}]{2019RNAAS...3..176O}
\bibinfo{author}{{Orlando}, S.}, \bibinfo{author}{{Pillitteri}, I.},
  \bibinfo{author}{{Bocchino}, F.}, \bibinfo{author}{{Daricello}, L.},
  \bibinfo{author}{{Leonardi}, L.}, \bibinfo{year}{2019}.
\newblock \bibinfo{title}{{3DMAP-VR, A Project to Visualize Three-dimensional
  Models of Astrophysical Phenomena in Virtual Reality}}.
\newblock \bibinfo{journal}{Research Notes of the American Astronomical
  Society} \bibinfo{volume}{3}, \bibinfo{pages}{176}.
\bibitem[{{Osinski} et~al.(2019){Osinski}, {Marion}, {Morse}, {Newman},
  {Pilles}, {Tornabene}, {Caudill} and {Christoffersen}}]{2019AGUFM.P33C..08O}
\bibinfo{author}{{Osinski}, G.}, \bibinfo{author}{{Marion}, C.},
  \bibinfo{author}{{Morse}, Z.}, \bibinfo{author}{{Newman}, J.},
  \bibinfo{author}{{Pilles}, E.}, \bibinfo{author}{{Tornabene}, L.L.},
  \bibinfo{author}{{Caudill}, C.}, \bibinfo{author}{{Christoffersen}, P.},
  \bibinfo{year}{2019}.
\newblock \bibinfo{title}{{Returning to the Moon: CanMoon and the Role of
  Analogue Missions}}, in: \bibinfo{booktitle}{AGU Fall Meeting Abstracts}, pp.
  \bibinfo{pages}{P33C--08}.
\bibitem[{{Pietriga} et~al.(2016){Pietriga}, {del Campo}, {Ibsen}, {Primet},
  {Appert}, {Chapuis}, {Hempel}, {Mu{\~n}oz}, {Eyheramendy}, {Jordan} and
  {Dole}}]{Pietriga16}
\bibinfo{author}{{Pietriga}, E.}, \bibinfo{author}{{del Campo}, F.},
  \bibinfo{author}{{Ibsen}, A.}, \bibinfo{author}{{Primet}, R.},
  \bibinfo{author}{{Appert}, C.}, \bibinfo{author}{{Chapuis}, O.},
  \bibinfo{author}{{Hempel}, M.}, \bibinfo{author}{{Mu{\~n}oz}, R.},
  \bibinfo{author}{{Eyheramendy}, S.}, \bibinfo{author}{{Jordan}, A.},
  \bibinfo{author}{{Dole}, H.}, \bibinfo{year}{2016}.
\newblock \bibinfo{title}{{Exploratory visualization of astronomical data on
  ultra-high-resolution wall displays}}, in: \bibinfo{editor}{{Chiozzi}, G.},
  \bibinfo{editor}{{Guzman}, J.C.} (Eds.), \bibinfo{booktitle}{Software and
  Cyberinfrastructure for Astronomy IV}, p. \bibinfo{pages}{99130W}.
\bibitem[{{Polsterer} and {Taylor}(2017)}]{2017ASPC..512..485P}
\bibinfo{author}{{Polsterer}, K.L.}, \bibinfo{author}{{Taylor}, M.B.},
  \bibinfo{year}{2017}.
\newblock \bibinfo{title}{{Virtual Observatory Virtual Reality}}, in:
  \bibinfo{editor}{{Lorente}, N.P.F.}, \bibinfo{editor}{{Shortridge}, K.},
  \bibinfo{editor}{{Wayth}, R.} (Eds.), \bibinfo{booktitle}{Astronomical Data
  Analysis Software and Systems XXV}, p. \bibinfo{pages}{485}.
\bibitem[{{Possami} et~al.(2021){Possami}, {Possami-Inesedy}, {Corpuz} and
  {Greenaway}}]{Possami21}
\bibinfo{author}{{Possami}, A.}, \bibinfo{author}{{Possami-Inesedy}, A.},
  \bibinfo{author}{{Corpuz}, G.}, \bibinfo{author}{{Greenaway}, E.},
  \bibinfo{year}{2021}.
\newblock \bibinfo{title}{{Got sick of surveys or lack of social capital? An
  investigation on the effects of the COVID-19 lockdown on institutional
  surveying}}.
\newblock \bibinfo{journal}{The Australian Educational Researcher}
  \bibinfo{volume}{51}, \bibinfo{pages}{21--39}.
\bibitem[{{Ram{\'\i}rez} et~al.(2019){Ram{\'\i}rez}, {Gonz{\'a}lez
  N{\'u}{\~n}ez}, {Hernandez}, {Salgado}, {Mora}, {Lammers}, {Mer{\'\i}n},
  {Baines}, {de Marchi} and {Arviset}}]{2019ASPC..523...21R}
\bibinfo{author}{{Ram{\'\i}rez}, E.}, \bibinfo{author}{{Gonz{\'a}lez
  N{\'u}{\~n}ez}, J.G.N.}, \bibinfo{author}{{Hernandez}, J.},
  \bibinfo{author}{{Salgado}, J.}, \bibinfo{author}{{Mora}, A.},
  \bibinfo{author}{{Lammers}, U.}, \bibinfo{author}{{Mer{\'\i}n}, B.},
  \bibinfo{author}{{Baines}, D.}, \bibinfo{author}{{de Marchi}, G.},
  \bibinfo{author}{{Arviset}, C.}, \bibinfo{year}{2019}.
\newblock \bibinfo{title}{{Analysis of Astronomical Data using VR: the Gaia
  Catalog in 3D}}, in: \bibinfo{editor}{{Teuben}, P.J.},
  \bibinfo{editor}{{Pound}, M.W.}, \bibinfo{editor}{{Thomas}, B.A.},
  \bibinfo{editor}{{Warner}, E.M.} (Eds.), \bibinfo{booktitle}{Astronomical
  Data Analysis Software and Systems XXVII}, p.~\bibinfo{pages}{21}.
\bibitem[{{Ramsey} et~al.(2024){Ramsey}, {Kastner}, {Butler}, {Binks} and
  {Skillman}}]{2024AAS...24345816R}
\bibinfo{author}{{Ramsey}, B.}, \bibinfo{author}{{Kastner}, J.},
  \bibinfo{author}{{Butler}, R.}, \bibinfo{author}{{Binks}, A.},
  \bibinfo{author}{{Skillman}, T.}, \bibinfo{year}{2024}.
\newblock \bibinfo{title}{{Exploring NGC 2287: Insights from Gaia and TESS}},
  in: \bibinfo{booktitle}{American Astronomical Society Meeting Abstracts}, p.
  \bibinfo{pages}{458.16}.
\bibitem[{{Romano} et~al.(2019){Romano}, {Burton}, {Ashley}, {Molinari},
  {Rebolledo}, {Braiding} and {Schisano}}]{2019MNRAS.484.2089R}
\bibinfo{author}{{Romano}, D.}, \bibinfo{author}{{Burton}, M.G.},
  \bibinfo{author}{{Ashley}, M.C.B.}, \bibinfo{author}{{Molinari}, S.},
  \bibinfo{author}{{Rebolledo}, D.}, \bibinfo{author}{{Braiding}, C.},
  \bibinfo{author}{{Schisano}, E.}, \bibinfo{year}{2019}.
\newblock \bibinfo{title}{{The G332 molecular cloud ring: I. Morphology and
  physical characteristics}}.
\newblock \bibinfo{journal}{MNRAS} \bibinfo{volume}{484},
  \bibinfo{pages}{2089--2118}.
\bibitem[{{Rosenfield} et~al.(2018){Rosenfield}, {Fay}, {Gilchrist}, {Cui},
  {Weigel}, {Robitaille}, {Otor} and {Goodman}}]{2018ApJS..236...22R}
\bibinfo{author}{{Rosenfield}, P.}, \bibinfo{author}{{Fay}, J.},
  \bibinfo{author}{{Gilchrist}, R.K.}, \bibinfo{author}{{Cui}, C.},
  \bibinfo{author}{{Weigel}, A.D.}, \bibinfo{author}{{Robitaille}, T.},
  \bibinfo{author}{{Otor}, O.J.}, \bibinfo{author}{{Goodman}, A.},
  \bibinfo{year}{2018}.
\newblock \bibinfo{title}{{AAS WorldWide Telescope: A Seamless, Cross-platform
  Data Visualization Engine for Astronomy Research, Education, and
  Democratizing Data}}.
\newblock \bibinfo{journal}{ApJS} \bibinfo{volume}{236}, \bibinfo{pages}{22}.
\bibitem[{{Rots}(1986)}]{Rots86}
\bibinfo{author}{{Rots}, A.}, \bibinfo{year}{1986}.
\newblock \bibinfo{title}{{Data display: searching for new avenues in image
  analysis.}}, in: \bibinfo{editor}{{Perley}, R.A.}, \bibinfo{editor}{{Schwab},
  F.R.}, \bibinfo{editor}{{Bridle}, A.H.} (Eds.), \bibinfo{booktitle}{Synthesis
  Imaging}, pp. \bibinfo{pages}{231--252}.
\bibitem[{{Russell} et~al.(2019){Russell}, {Sepulveda}, {Cuadra} and
  {Wang}}]{2019HEAD...1711286R}
\bibinfo{author}{{Russell}, C.M.P.}, \bibinfo{author}{{Sepulveda}, M.},
  \bibinfo{author}{{Cuadra}, J.}, \bibinfo{author}{{Wang}, Q.},
  \bibinfo{year}{2019}.
\newblock \bibinfo{title}{{A hundred stellar winds, some X-rays, and Sgr A*
  walk into VR{\textellipsis}}}, in: \bibinfo{booktitle}{AAS/High Energy
  Astrophysics Division}, p. \bibinfo{pages}{112.86}.
\bibitem[{{Russell} et~al.(2018){Russell}, {Wang} and
  {Cuadra}}]{2018cosp...42E2915R}
\bibinfo{author}{{Russell}, C.M.P.}, \bibinfo{author}{{Wang}, Q.D.},
  \bibinfo{author}{{Cuadra}, J.}, \bibinfo{year}{2018}.
\newblock \bibinfo{title}{{The inner parsec of the Galactic center:
  hydrodynamics, X-ray modeling, and 360-degree videos}}, in:
  \bibinfo{booktitle}{42nd COSPAR Scientific Assembly}, pp.
  \bibinfo{pages}{E1.11--11--18}.
\bibitem[{{Savani} et~al.(2018){Savani}, {Boyer} and
  {Olano}}]{2018shin.confE.194S}
\bibinfo{author}{{Savani}, N.P.}, \bibinfo{author}{{Boyer}, K.},
  \bibinfo{author}{{Olano}, M.}, \bibinfo{year}{2018}.
\newblock \bibinfo{title}{{Scientific insights from visualizing 3D simulations
  in Virtual Reality environment for PSP}}, in: \bibinfo{booktitle}{Solar
  Heliospheric and INterplanetary Environment (SHINE 2018)}, p.
  \bibinfo{pages}{194}.
\bibitem[{{Schaaff} et~al.(2015){Schaaff}, {Berthier}, {Da Rocha}, {Deparis},
  {Derriere}, {Gaultier}, {Houpin}, {Normand} and {Ocvirk}}]{Schaaff15}
\bibinfo{author}{{Schaaff}, A.}, \bibinfo{author}{{Berthier}, J.},
  \bibinfo{author}{{Da Rocha}, J.}, \bibinfo{author}{{Deparis}, N.},
  \bibinfo{author}{{Derriere}, S.}, \bibinfo{author}{{Gaultier}, P.},
  \bibinfo{author}{{Houpin}, R.}, \bibinfo{author}{{Normand}, J.},
  \bibinfo{author}{{Ocvirk}, P.}, \bibinfo{year}{2015}.
\newblock \bibinfo{title}{{Immersive 3D Visualization of Astronomical Data}},
  in: \bibinfo{editor}{{Taylor}, A.R.}, \bibinfo{editor}{{Rosolowsky}, E.}
  (Eds.), \bibinfo{booktitle}{Astronomical Data Analysis Software an Systems
  XXIV (ADASS XXIV)}, p. \bibinfo{pages}{125}.
\bibitem[{{Sen} et~al.(2022){Sen}, {Agarwal}, {Chakraborty} and
  {Singh}}]{Sen22}
\bibinfo{author}{{Sen}, S.}, \bibinfo{author}{{Agarwal}, S.},
  \bibinfo{author}{{Chakraborty}, P.}, \bibinfo{author}{{Singh}, K.P.},
  \bibinfo{year}{2022}.
\newblock \bibinfo{title}{{Astronomical big data processing using machine
  learning: A comprehensive review}}.
\newblock \bibinfo{journal}{Experimental Astronomy} \bibinfo{volume}{53},
  \bibinfo{pages}{1--43}.
\bibitem[{{Serra} et~al.(2015){Serra}, {Westmeier}, {Giese}, {Jurek},
  {Fl{\"o}er}, {Popping}, {Winkel}, {van der Hulst}, {Meyer}, {Koribalski},
  {Staveley-Smith} and {Courtois}}]{Serra15}
\bibinfo{author}{{Serra}, P.}, \bibinfo{author}{{Westmeier}, T.},
  \bibinfo{author}{{Giese}, N.}, \bibinfo{author}{{Jurek}, R.},
  \bibinfo{author}{{Fl{\"o}er}, L.}, \bibinfo{author}{{Popping}, A.},
  \bibinfo{author}{{Winkel}, B.}, \bibinfo{author}{{van der Hulst}, T.},
  \bibinfo{author}{{Meyer}, M.}, \bibinfo{author}{{Koribalski}, B.S.},
  \bibinfo{author}{{Staveley-Smith}, L.}, \bibinfo{author}{{Courtois}, H.},
  \bibinfo{year}{2015}.
\newblock \bibinfo{title}{{SOFIA: a flexible source finder for 3D spectral line
  data}}.
\newblock \bibinfo{journal}{MNRAS} \bibinfo{volume}{448},
  \bibinfo{pages}{1922--1929}.
\bibitem[{{Severson} et~al.(2020){Severson}, {Kassis}, {Windmiller}, {SSU VR
  Team} and {SDSU VR Team}}]{2020AAS...23520204S}
\bibinfo{author}{{Severson}, S.}, \bibinfo{author}{{Kassis}, S.},
  \bibinfo{author}{{Windmiller}, G.}, \bibinfo{author}{{SSU VR Team}},
  \bibinfo{author}{{SDSU VR Team}}, \bibinfo{year}{2020}.
\newblock \bibinfo{title}{{Evaluating Virtual Reality as a Tool for Astronomy
  Education}}, in: \bibinfo{booktitle}{American Astronomical Society Meeting
  Abstracts}, p. \bibinfo{pages}{202.04}.
\bibitem[{{Sims} et~al.(2010){Sims}, {Dodson} and {Edwards}}]{Sims10}
\bibinfo{author}{{Sims}, M.H.}, \bibinfo{author}{{Dodson}, K.E.},
  \bibinfo{author}{{Edwards}, L.J.}, \bibinfo{year}{2010}.
\newblock \bibinfo{title}{{Hyperwall Use as a Tool for Collaboration}}, in:
  \bibinfo{booktitle}{Astrobiology Science Conference 2010: Evolution and Life:
  Surviving Catastrophes and Extremes on Earth and Beyond}, p.
  \bibinfo{pages}{5614}.
\bibitem[{{Toussaint} et~al.(2020){Toussaint}, {Koelemeijer} and
  {Zaroli}}]{2020EGUGA..2211708T}
\bibinfo{author}{{Toussaint}, R.}, \bibinfo{author}{{Koelemeijer}, P.},
  \bibinfo{author}{{Zaroli}, C.}, \bibinfo{year}{2020}.
\newblock \bibinfo{title}{{Inside blue dots - Grasping dynamic global fields
  thanks to Virtual Reality}}, in: \bibinfo{booktitle}{EGU General Assembly
  Conference Abstracts}, p. \bibinfo{pages}{11708}.
\bibitem[{{Virtanen} et~al.(2020){Virtanen}, {Gommers}, {Oliphant},
  {Haberland}, {Reddy}, {Cournapeau}, {Burovski}, {Peterson}, {Weckesser},
  {Bright}, {van der Walt}, {Brett}, {Wilson}, {Millman}, {Mayorov}, {Nelson},
  {Jones}, {Kern}, {Larson}, {Carey}, {Polat}, {Feng}, {Moore}, {VanderPlas},
  {Laxalde}, {Perktold}, {Cimrman}, {Henriksen}, {Quintero}, {Harris},
  {Archibald}, {Ribeiro}, {Pedregosa}, {van Mulbregt} and {SciPy 1. 0
  Contributors}}]{2020NatMe..17..261V}
\bibinfo{author}{{Virtanen}, P.}, \bibinfo{author}{{Gommers}, R.},
  \bibinfo{author}{{Oliphant}, T.E.}, \bibinfo{author}{{Haberland}, M.},
  \bibinfo{author}{{Reddy}, T.}, \bibinfo{author}{{Cournapeau}, D.},
  \bibinfo{author}{{Burovski}, E.}, \bibinfo{author}{{Peterson}, P.},
  \bibinfo{author}{{Weckesser}, W.}, \bibinfo{author}{{Bright}, J.},
  \bibinfo{author}{{van der Walt}, S.J.}, \bibinfo{author}{{Brett}, M.},
  \bibinfo{author}{{Wilson}, J.}, \bibinfo{author}{{Millman}, K.J.},
  \bibinfo{author}{{Mayorov}, N.}, \bibinfo{author}{{Nelson}, A.R.J.},
  \bibinfo{author}{{Jones}, E.}, \bibinfo{author}{{Kern}, R.},
  \bibinfo{author}{{Larson}, E.}, \bibinfo{author}{{Carey}, C.J.},
  \bibinfo{author}{{Polat}, {\.I}.}, \bibinfo{author}{{Feng}, Y.},
  \bibinfo{author}{{Moore}, E.W.}, \bibinfo{author}{{VanderPlas}, J.},
  \bibinfo{author}{{Laxalde}, D.}, \bibinfo{author}{{Perktold}, J.},
  \bibinfo{author}{{Cimrman}, R.}, \bibinfo{author}{{Henriksen}, I.},
  \bibinfo{author}{{Quintero}, E.A.}, \bibinfo{author}{{Harris}, C.R.},
  \bibinfo{author}{{Archibald}, A.M.}, \bibinfo{author}{{Ribeiro}, A.H.},
  \bibinfo{author}{{Pedregosa}, F.}, \bibinfo{author}{{van Mulbregt}, P.},
  \bibinfo{author}{{SciPy 1. 0 Contributors}}, \bibinfo{year}{2020}.
\newblock \bibinfo{title}{{SciPy 1.0: fundamental algorithms for scientific
  computing in Python}}.
\newblock \bibinfo{journal}{Nature Methods} \bibinfo{volume}{17},
  \bibinfo{pages}{261--272}.
\bibitem[{{Vizi} and {B{\'e}rczi}(2020)}]{2020LPI....51.2916V}
\bibinfo{author}{{Vizi}, P.G.}, \bibinfo{author}{{B{\'e}rczi}, S.},
  \bibinfo{year}{2020}.
\newblock \bibinfo{title}{{Apollo Memorial Year for the Planetary Science
  Education - Real and VR Exhibition}}, in: \bibinfo{booktitle}{Lunar and
  Planetary Science Conference}, p. \bibinfo{pages}{2916}.
\bibitem[{{Walker} et~al.(2018){Walker}, {Burns} and
  {Szafir}}]{2018LPICo2063.3095W}
\bibinfo{author}{{Walker}, M.E.}, \bibinfo{author}{{Burns}, J.O.},
  \bibinfo{author}{{Szafir}, D.J.}, \bibinfo{year}{2018}.
\newblock \bibinfo{title}{{VR Simulation Testbed: Improving Surface
  Telerobotics for the Deep Space Gateway}}, in: \bibinfo{booktitle}{Deep Space
  Gateway Concept Science Workshop}, p. \bibinfo{pages}{3095}.
\bibitem[{{Webb} and {Goode}(2023)}]{Webb2023}
\bibinfo{author}{{Webb}, S.A.}, \bibinfo{author}{{Goode}, S.R.},
  \bibinfo{year}{2023}.
\newblock \bibinfo{title}{{An Astronomers Guide to Machine Learning}}.
\newblock \bibinfo{journal}{arXiv e-prints}
  \href{http://arxiv.org/abs/2304.00512}{\tt arXiv:2304.00512}.
\bibitem[{{Westmeier} et~al.(2022){Westmeier}, {Deg}, {Spekkens}, {Reynolds},
  {Shen}, {Gaudet}, {Goliath}, {Huynh}, {Venkataraman}, {Lin}, {O'Beirne},
  {Catinella}, {Cortese}, {D{\'e}nes}, {Elagali}, {For}, {J{\'o}zsa},
  {Howlett}, {van der Hulst}, {Jurek}, {Kamphuis}, {Kilborn}, {Kleiner},
  {Koribalski}, {Lee-Waddell}, {Murugeshan}, {Rhee}, {Serra}, {Shao},
  {Staveley-Smith}, {Wang}, {Wong}, {Zwaan}, {Allison}, {Anderson}, {Ball},
  {Bock}, {Brodrick}, {Bunton}, {Cooray}, {Gupta}, {Hayman}, {Mahony}, {Moss},
  {Ng}, {Pearce}, {Raja}, {Roxby}, {Voronkov}, {Warhurst}, {Courtois} and
  {Said}}]{Westmeier22}
\bibinfo{author}{{Westmeier}, T.}, \bibinfo{author}{{Deg}, N.},
  \bibinfo{author}{{Spekkens}, K.}, \bibinfo{author}{{Reynolds}, T.N.},
  \bibinfo{author}{{Shen}, A.X.}, \bibinfo{author}{{Gaudet}, S.},
  \bibinfo{author}{{Goliath}, S.}, \bibinfo{author}{{Huynh}, M.T.},
  \bibinfo{author}{{Venkataraman}, P.}, \bibinfo{author}{{Lin}, X.},
  \bibinfo{author}{{O'Beirne}, T.}, \bibinfo{author}{{Catinella}, B.},
  \bibinfo{author}{{Cortese}, L.}, \bibinfo{author}{{D{\'e}nes}, H.},
  \bibinfo{author}{{Elagali}, A.}, \bibinfo{author}{{For}, B.Q.},
  \bibinfo{author}{{J{\'o}zsa}, G.I.G.}, \bibinfo{author}{{Howlett}, C.},
  \bibinfo{author}{{van der Hulst}, J.M.}, \bibinfo{author}{{Jurek}, R.J.},
  \bibinfo{author}{{Kamphuis}, P.}, \bibinfo{author}{{Kilborn}, V.A.},
  \bibinfo{author}{{Kleiner}, D.}, \bibinfo{author}{{Koribalski}, B.S.},
  \bibinfo{author}{{Lee-Waddell}, K.}, \bibinfo{author}{{Murugeshan}, C.},
  \bibinfo{author}{{Rhee}, J.}, \bibinfo{author}{{Serra}, P.},
  \bibinfo{author}{{Shao}, L.}, \bibinfo{author}{{Staveley-Smith}, L.},
  \bibinfo{author}{{Wang}, J.}, \bibinfo{author}{{Wong}, O.I.},
  \bibinfo{author}{{Zwaan}, M.A.}, \bibinfo{author}{{Allison}, J.R.},
  \bibinfo{author}{{Anderson}, C.S.}, \bibinfo{author}{{Ball}, L.},
  \bibinfo{author}{{Bock}, D.C.J.}, \bibinfo{author}{{Brodrick}, D.},
  \bibinfo{author}{{Bunton}, J.D.}, \bibinfo{author}{{Cooray}, F.R.},
  \bibinfo{author}{{Gupta}, N.}, \bibinfo{author}{{Hayman}, D.B.},
  \bibinfo{author}{{Mahony}, E.K.}, \bibinfo{author}{{Moss}, V.A.},
  \bibinfo{author}{{Ng}, A.}, \bibinfo{author}{{Pearce}, S.E.},
  \bibinfo{author}{{Raja}, W.}, \bibinfo{author}{{Roxby}, D.N.},
  \bibinfo{author}{{Voronkov}, M.A.}, \bibinfo{author}{{Warhurst}, K.A.},
  \bibinfo{author}{{Courtois}, H.M.}, \bibinfo{author}{{Said}, K.},
  \bibinfo{year}{2022}.
\newblock \bibinfo{title}{{WALLABY pilot survey: Public release of H I data for
  almost 600 galaxies from phase 1 of ASKAP pilot observations}}.
\newblock \bibinfo{journal}{PASA} \bibinfo{volume}{39}, \bibinfo{pages}{e058}.
\bibitem[{{Westmeier} et~al.(2021){Westmeier}, {Kitaeff}, {Pallot}, {Serra},
  {van der Hulst}, {Jurek}, {Elagali}, {For}, {Kleiner}, {Koribalski},
  {Lee-Waddell}, {Mould}, {Reynolds}, {Rhee} and
  {Staveley-Smith}}]{Westmeier21}
\bibinfo{author}{{Westmeier}, T.}, \bibinfo{author}{{Kitaeff}, S.},
  \bibinfo{author}{{Pallot}, D.}, \bibinfo{author}{{Serra}, P.},
  \bibinfo{author}{{van der Hulst}, J.M.}, \bibinfo{author}{{Jurek}, R.J.},
  \bibinfo{author}{{Elagali}, A.}, \bibinfo{author}{{For}, B.Q.},
  \bibinfo{author}{{Kleiner}, D.}, \bibinfo{author}{{Koribalski}, B.S.},
  \bibinfo{author}{{Lee-Waddell}, K.}, \bibinfo{author}{{Mould}, J.R.},
  \bibinfo{author}{{Reynolds}, T.N.}, \bibinfo{author}{{Rhee}, J.},
  \bibinfo{author}{{Staveley-Smith}, L.}, \bibinfo{year}{2021}.
\newblock \bibinfo{title}{{SOFIA 2 - an automated, parallel H I source finding
  pipeline for the WALLABY survey}}.
\newblock \bibinfo{journal}{MNRAS} \bibinfo{volume}{506},
  \bibinfo{pages}{3962--3976}.
\bibitem[{{Wiedemann} et~al.(2020){Wiedemann}, {Schuberth}, {Colli}, {Bunge}
  and {Kranzlm{\"u}ller}}]{2020EGUGA..22.5714W}
\bibinfo{author}{{Wiedemann}, M.}, \bibinfo{author}{{Schuberth}, B.S.A.},
  \bibinfo{author}{{Colli}, L.}, \bibinfo{author}{{Bunge}, H.P.},
  \bibinfo{author}{{Kranzlm{\"u}ller}, D.}, \bibinfo{year}{2020}.
\newblock \bibinfo{title}{{Visualising large-scale geodynamic simulations: How
  to Dive into Earth's Mantle with Virtual Reality}}, in:
  \bibinfo{booktitle}{EGU General Assembly Conference Abstracts}, p.
  \bibinfo{pages}{5714}.
\bibitem[{{Zelinka} et~al.(2021){Zelinka}, {Truong}, {Bao}, {Kojecky} and
  {Amer}}]{Zelink21}
\bibinfo{author}{{Zelinka}, I.}, \bibinfo{author}{{Truong}, T.C.},
  \bibinfo{author}{{Bao}, D.Q.}, \bibinfo{author}{{Kojecky}, L.},
  \bibinfo{author}{{Amer}, E.}, \bibinfo{year}{2021}.
\newblock \bibinfo{title}{{Artifical Intelligence in Astrophysics}}, in:
  \bibinfo{editor}{{Zelinka}, I.}, \bibinfo{editor}{{Brescia}, M.},
  \bibinfo{editor}{{Baron}, D.} (Eds.), \bibinfo{booktitle}{Intelligent
  Astrophysics. Edited by I. Zelinka}, pp. \bibinfo{pages}{1--28}.

\end{thebibliography}

\appendix
\section{Population proportion confidence intervals}
\label{pci}
For each of the 10 questions in the AIDA2 survey, we calculate the population proportion confidence interval using the procedure below.  For a stated confidence level, the confidence interval provides an estimate of the uncertainty that a measured value represents the population.

Given that a particular answer was provided $N_{\rm response}$ times -- for example, $N_{\rm response} = 16$ selected ``Yes'' to question Q1 {\em Have you used a standard image display?} -- we calculate the sample proportion to be:
\begin{equation}
P_{\rm sp} = \frac{N_{\rm response}}{N_{\rm sample}}.
\end{equation}

In all cases, the sample size of respondents is $N_{\rm sample} = 17$ from a population of $N_{\rm population} =750$ members of the ASA. From this we determine the finite population correction:
\begin{equation}
FPC = \frac{N_{\rm population} - N_{\rm sample}}{N_{\rm population}-1} = 0.9786
\end{equation}

Selecting the required confidence level, the population proportion confidence interval (CI) is then:
\begin{equation}
CI = P_{\rm sp} \pm Z_{a/2} \sqrt{FPC \times \frac{P_{\rm sp}(1-P_{\rm sp})}{N_{\rm sample}}}
\end{equation}
where $Z_{a/2}$ is the critical value of the normal distribution. For a confidence level of $90\%$, $a = 0.1$ and $Z_{a/2} = 1.65$ using a standard normal table.
\end{document}